\newif\ifappendix\appendixtrue
\newif\ifmonitor\monitorfalse

\pdfoutput=1

\documentclass[a4paper,english,11pt]{scrartcl}
\usepackage{amsthm}
\usepackage{amsmath}
\usepackage{amssymb}
\usepackage{mathtools}
\usepackage[bookmarks=false,colorlinks=true, linkcolor=blue, urlcolor=blue, citecolor=blue, breaklinks=true]{hyperref}
\usepackage{cleveref}
\usepackage[natbib]{biblatex}
\date{}

\ifmonitor\usepackage[width=160mm, height=247mm,center]{crop}\fi

\usepackage{amssymb}
\usepackage{mathtools}
\usepackage{longtable}
\usepackage[caption=false]{subfig}

\usepackage{tikz-cd} 

\crefname{cons}{constraint}{constraints}
\Crefname{cons}{Constraint}{Constraints}
\creflabelformat{cons}{(#2#1#3)}

\crefname{claim}{claim}{claims}
\Crefname{claim}{Claim}{Claims}
\crefrangeformat{prop}{propositions~#3#1#4 to~#5#2#6}
\Crefrangeformat{prop}{Propositions~#3#1#4 to~#5#2#6}
\crefmultiformat{prop}{propositions~#2#1#3}{ and~#2#1#3}{, #2#1#3}{ and~#2#1#3}
\crefname{subsection}{section}{sections}
\Crefname{subsection}{Section}{Sections}

\usepackage{braket}					\usepackage{bm}						\usepackage{nicefrac}

\usepackage{tikz}
\usepackage{wrapfig}
\usepackage{adjustbox}
\usepackage{mathrsfs, stmaryrd}

\usepackage{enumitem}

\usetikzlibrary{chains,shapes.multipart}
\usetikzlibrary{shapes,calc}
\usetikzlibrary{automata,positioning}
\usetikzlibrary{arrows, decorations.markings}
\usetikzlibrary{shapes}

\usepackage{pgfplots}
\pgfplotsset{compat=1.17} 
\usepgfplotslibrary{statistics} 

\tikzstyle{vertex} = [shape=circle,draw=black]
\tikzstyle{namedVertex} = [shape=circle,draw=black]
\tikzstyle{edge} = [draw,->,thick]
\tikzstyle{labeledNodeS}=[circle, color=black!75!white, draw, inner sep = 0.1em, minimum size = 1.5em, scale=1.25]
\tikzstyle{normalEdge}=[very thick, >=stealth]
\tikzstyle{labeledNode}=[circle, draw, inner sep = 0.1em, minimum size = 1.5em, scale=0.75, fill=white]
\tikzstyle{normalEdgeF}=[line width=1.6pt, color=black!75!white, >=stealth]

\newtheorem{theorem}{Theorem}
\newtheorem{lemma}[theorem]{Lemma}
\newtheorem{proposition}[theorem]{Proposition}
\newtheorem{corollary}[theorem]{Corollary}
\newtheorem{claim}[theorem]{Claim}

\theoremstyle{definition}
\newtheorem{definition}[theorem]{Definition}
\newtheorem{example}[theorem]{Example}
\theoremstyle{remark}
\newtheorem{remark}[theorem]{Remark}
\newtheorem{notation}[theorem]{Notation}

\newenvironment{proofClaim}[1][]{\ifthenelse{\equal{#1}{}}{\begin{proof}}{\begin{proof}[#1]}}{\end{proof}}

\title{Machine-Learned Prediction Equilibrium for Dynamic Traffic Assignment\footnote{This version of the paper differs rather significantly from its initial arXiv submission:
While the first version represents the full version of the conference paper~\cite{Graf_Harks_Kollias_Markl_2022}, this updated version contains numerous generalizations that were introduced in the journal version~\cite{JMLR:v24:22-1446}.}}

\usepackage{authblk}
\author[1]{Lukas Graf}
\author[1]{Tobias Harks}
\author[2]{Kostas Kollias}
\author[1]{Michael Markl}
\affil[1]{University of Augsburg}
\affil[2]{Google}
\affil[ ]{\normalsize\href{mailto:lukas.graf@uni-passau.de;\%20tobias.harks@uni-passau.de;\%20kostaskollias@google.com;\%20michael.markl@uni-passau.de?subject=Comment\%20on\%20your\%20paper\%20\%22Machine-Learned\%20Prediction\%20Equilibrium\%20for\%20Dynamic\%20Traffic\%20Assignment\%22}{\{lukas.graf, tobias.harks\}@uni-passau.de, kostaskollias@google.com, michael.markl@uni-passau.de}}

\usepackage{xcolor}
\usepackage{soul}
\usepackage[normalem]{ulem}

\definecolor{darkgreen}{rgb}{0.0,0.9,0.0}

\newcommand{\R}{\mathbb R}
\newcommand{\N}{\mathbb N}

\newcommand{\predZ}{{\mathrm{Z}}}
\newcommand{\predC}{{\mathrm{C}}}
\newcommand{\predL}{{\mathrm{L}}}
\newcommand{\predLR}{{\mathrm{LR}}}
\newcommand{\predNN}{{\mathrm{NN}}}
\newcommand{\predRL}{{\mathrm{RL}}}
\newcommand{\predMLRaw}{{\mathrm{ML,raw}}}
\newcommand{\predML}{{\mathrm{ML}}}
\newcommand{\predP}{{\mathrm{P}}}

\newcommand{\predq}{\hat q}
\newcommand{\predE}{\hat E}
\newcommand{\predT}{\hat T}
\newcommand{\predTP}[2]{\hat\ell_{#1}^{#2}}
\newcommand{\predl}{\hat\ell}
\newcommand{\predDel}{\hat \Delta}
\newcommand{\transit}{\tau}
\newcommand{\mintransit}{\transit_{\min}}
\newcommand{\maxoutdeg}{\mathrm{d}^+_{\max}}

\newcommand{\capa}{\nu}
\newcommand{\comp}{\circ}
\newcommand{\paths}{\mathcal P}
\newcommand{\Lloc}[1]{L_{\mathrm{loc}}^{#1}}
\newcommand{\aequiv}{\overset{a.e.}{\equiv}}
\newcommand{\aequal}{\overset{a.e.}{=}}
\newcommand{\aleq}{\overset{a.e.}{\leq}}
\newcommand{\restrict}[2]{{#1}\vert_{#2}}
\newcommand{\restrictUpTo}[2]{#1_{\leq #2}}
\newcommand{\RateFcts}{\mathcal{R}}
\newcommand{\emptyArg}{\,\,\boldsymbol\cdot\,\,}
\newcommand{\firstN}[1]{[#1]}
\newcommand{\leb}{\lambda}
\newcommand{\bigO}{\mathcal{O}}

\newcommand{\outEdges}[1]{\delta^+_{#1}}

\newcommand{\inEdges}[1]{\delta^-_{#1}}

\newcommand{\infl}[1][f]{{#1}^+}
\newcommand{\outfl}[1][f]{{#1}^-}

\newcommand{\floor}[1]{\left\lfloor {#1} \right\rfloor}
\newcommand{\ceil}[1]{\left\lceil {#1} \right\rceil}
\newcommand{\abs}[1]{\left| #1 \right|}
\newcommand{\smallabs}[1]{| #1 |}
\newcommand{\norm}[1]{\left\lVert #1 \right\rVert}
\newcommand{\smallnorm}[1]{\lVert #1 \rVert}

\newcommand{\colonequiv}{\vcentcolon\mspace{-1.2mu}\equiv}

\usepackage{dsfont}
\newcommand{\one}{\mathds{1}}
\newcommand{\CharF}[1]{\one_{#1}}
\newcommand{\Hcomp}{H_{\mathrm{comp}}}

\DeclarePairedDelimiterX{\scalar}[2]{\langle}{\rangle}{#1, #2}
\DeclarePairedDelimiterX{\scalprod}[2]{\langle}{\rangle}{#1, #2}

\newcommand{\edgesLeaving}[1]{\delta^+_{#1}}
\newcommand{\edgesEntering}[1]{\delta^-_{#1}}

\newcommand*\diff{\mathop{}\!\mathrm{d}}

\newcommand{\depl}{\mathrm{depl}}

\newcommand{\piecewise}{piecewise}

\newcommand{\displayStyleForArxiv}[1]{
\[#1\]
}

\begin{document}

\maketitle

\begin{abstract}
We study a dynamic traffic assignment model, where agents base their instantaneous routing decisions on real-time delay predictions.
We formulate a mathematically concise model and define {\em dynamic prediction equilibrium (DPE)} in which no agent can at any point during their journey improve their predicted travel time by switching to a different route.
We demonstrate the versatility of our framework by showing that it subsumes the well-known full information and instantaneous information models, in addition to admitting further realistic predictors as special cases.
We then proceed to derive properties of the predictors that ensure a dynamic prediction equilibrium exists.  
Additionally, we define \emph{$\varepsilon$-approximate DPE} wherein no agent can improve their predicted travel time by more than $\varepsilon$ and provide further conditions of the predictors under which such an approximate equilibrium can be computed.
Finally, we complement our theoretical analysis by an experimental study, in which we systematically compare the induced average travel times of different predictors, including two machine-learning based models trained on data gained from previously computed approximate equilibrium flows, both on synthetic and real world road networks. 
\end{abstract}

\clearpage

\tableofcontents

\clearpage

\section{Introduction}

Modelling and optimizing traffic flows is a significant effort that impacts billions of people living in urban areas, with key challenges including managing congestion and carbon emissions. These phenomena are heavily impacted by individual driver routing decisions, which are often influenced by ML-based predictions for the delays of road segments (see, for instance, \cite{GNNSurvey} for an overview of convolutional and graph neural network based approaches). One key aspect that is not well understood, is that these routing decisions, in turn, influence the forecasting models by changing the underlying signature of traffic flows
and thus lead to a complex and self-referential system.

In this paper, we address this interplay focusing on the popular dynamic traffic assignment (DTA) framework, on which there has been substantial work over the past decades (see the classical book of Ford and Fulkerson~\citealp{Ford62}, or the more recent surveys of Friesz et al.~\citealp{Friesz19}, Peeta and Ziliaskopoulos~\citealp{Peeta01}, and Skutella~\citealp{Skutella08}).
A fundamental base model describing the dynamic flow propagation process is the so-called \emph{deterministic queuing model} due to Vickrey~\cite{Vickrey69}.
Here, a directed graph $G=(V,E)$ is given, where edges
$e\in E$ are associated with a queue with a positive rate capacity $\nu_e$ and a physical transit time $\tau_e$.
If the total inflow into an edge  $e=vw\in E$ exceeds the rate capacity
$\nu_e$, a queue builds up and agents need to 
wait in the queue before they are forwarded along the edge. The total travel time
along $e$  is thus composed of the waiting time spent in the queue plus the physical transit time $\tau_e$.
The Vickrey model is arguably one of the most important traffic models  (see Li, Huang and Yang~\citealp{LI2020}, for an up to date research overview of the past 50 years),
and yet, it is mathematically quite challenging to analyze (see Friesz et al.~\citealp{Han2013a}, for a discussion of the inherent complexities).

Given a physical flow propagation model, the routing
and traffic prediction algorithms are usually subsumed under a \emph{behavioral model} of agents 
in order to solve a DTA model. The behavior of agents is  modelled based on  their informational assumption which in turn defines their respective utility function.
 Most works in the DTA literature on the Vickrey model can roughly be classified into
two main informational categories: the \emph{full information model} and the \emph{instantaneous information model}. In the full information model, an agent is able to exactly forecast future travel times
along a chosen path effectively anticipating the
whole spatio-temporal flow evolution over the network.
This assumption has been justified by letting travelers
learn good routes over several trips and
a dynamic equilibrium then corresponds to an attractor of an underlying learning dynamic.
Existence and computation of dynamic equilibria in the full information model have been studied extensively in the transportation science literature, see \cite{Friesz93,Han2013,Han2013a,Han2013b,MeunierW10,ZhuM00}, whereas the works by \cite{Koch11,CominettiCL15} allow a direct combinatorial characterization of dynamic equilibria leading to existence and uniqueness results in the realm of the Vickrey bottleneck model.  While certainly relevant and key for the entire development of the research in DTA, this concept may not  accurately reflect the behavioral changes caused by the wide-spread use of navigation devices
and resulting real-time decisions by agents. 

In the instantaneous route choice model, agents are informed in real-time about the current traffic situations and, if beneficial, reroute instantaneously no matter how good or bad that route was in hindsight, see Ran and Boyce~\cite[\S~VII-IX]{Ran96}, Boyce, Ran and LeBlanc~\cite[]{BoyceRL95,RanBL93}, Friesz et al.~\cite{FrieszLTW89}.
Indeed it seems more realistic that the information available to a navigation device is rather instantaneous and certainly not complete, that is, congestion information is available only as an aggregate (estimated waiting times for road traversal) but the individual routes and/or source and destination of other travelers are usually unknown.
For the Vickrey model, Graf, Harks and Sering~\cite{GrafHS20} established the existence of instantaneous dynamic equilibria and derived further structural properties.
A key difference between dynamic equilibria (in the full information model) and instantaneous dynamic equilibria is the possibility of \emph{cyclic behavior} in the latter.
More specifically,~\cite{GrafHS20} describe instances with only two origin-destination pairs and a finite flow volume in which \emph{any} instantaneous  dynamic equilibrium cycles forever.
This can never happen in the full information model as an agent plays a best-response given the collective decisions of all other agents, thus, any cycle only increases the travel time.

\subsection{Our Contribution}
We propose a new DTA formulation within the Vickrey model that is based on \emph{predicted} travel times.
Since the physical transit times are known a priori, the only unknown is the precise evolution of the flow over time.
In our model, agents use a \emph{queue prediction function} that provides for any future point in time a prediction of queues.
This model includes as special cases the full information model and the instantaneous information model but also allows the use of predictions based on historical data or the flow evolution learned en route.
The model also includes the case of finitely many \emph{classes} of agents similar to the concept described by~\cite{Dafermos72}, for which agents of the same class use the same prediction function, but the prediction function may vary across the different classes.
Technically, this is solved by subdividing the traffic commodities by their different classes and assigning a prediction function to each (sub-)commodity.

\paragraph{Existence Results.}
As our main theoretical contribution, we define this model formally 
and derive conditions for the queue predictors  leading to the existence of
dynamic (prediction) equilibria.
The main approach is based on an \emph{extension property} of partial equilibrium flows, that is, we show that any equilibrium flow up to some time $\theta\geq 0$ can be extended to time $\theta+\alpha$ for some fixed constant $\alpha>0$ which leads
to the existence on the whole~$\R$ after a countably infinite number of extensions.
The extension step itself is based on a formulation using infinite dimensional variational inequalities in the edge-flow space.
If the predictor satisfies a continuity condition, only depends on past information, and respects a First-In-First-Out (FIFO) rule, the extension is possible and leads to existence of an equilibrium.

While this approach is in line with previous existence proofs using
variational inequalities as put forth in the seminal papers by Friesz et al.~\cite{Friesz93,Han2013,Han2013a,Han2013b}, there are some remarkable differences.
The above works rely on the complete spatio-temporal unfolding of the path-inflows over the network which is known as \emph{network loading}.
As shown by~\cite{GrafHS20}, already the simple prediction function given by the constant current queues (which leads to the instantaneous route choice model) leads to dynamic equilibria with cycling behavior (forever) and thus puts a path-based formulation over the entire time horizon out of reach.
Our approach uses an extension-methodology, which does
not rely on the complete spatio-temporal unfolding of flow.
We demonstrate the applicability of our main result by showing that it applies not only to several simple variants of predictors but also to predictors that implement an arbitrary continuous (machine-learned) transformation.

\paragraph{Computing Approximate Dynamic Prediction Equilibria.} 
We show that under mild conditions on the predictors, an approximate version of a dynamic prediction equilibrium can be computed in finite time by introducing an extension-based algorithm.
This algorithm computes a flow in which agents recompute their predicted shortest path to the sink with the latest predictions in a periodic interval.

If the predictors behave according to a Lipschitz condition, we show that these flow instances approximate dynamic prediction flows.

\paragraph{Computational Study.}
On the experimental side, we conduct experiments on small synthetic networks, on the Sioux Falls network from \cite{LeBlanc1975}, and on a larger real road network of Anaheim obtained from~\cite{transportationnetworks}.
For various predictors, we study how the particles’ choice of the predictor affects their average travel time.
For this purpose, we also train linear regression and simple neural network models, for use as one of our predictors.

\paragraph{Comparison to Conference Version.}
An extended abstract of this paper was published in~\cite{Graf_Harks_Kollias_Markl_2022}.
In comparison, this work not only includes fully worked-out proofs of all theorems, but also generalizes the theoretical model of equilibria by making the predictors depend on the historical flow instead of its corresponding queue functions.
Appropriate and thus weaker conditions for the existence of equilibria are worked out such that the original result occurs as a special case.
This allows us to embed arbitrary continuous ML methods as opposed to only simple linear regression models.
A detailed description of the algorithm for computing approximate equilibria, and the proof of its correctness and termination are added.
Finally, we revised the computational study:
We implement a new neural network based predictor and evaluate all predictors in a larger experiment setup based on real world origin-destination demand pairs.

\subsection{Related Work}
The idea of using real-time information and traffic predictions en route and subsequently change the route is by no means new and has been proposed under varying names such as ATIS (advanced traveller information systems), see~\citep{Chosus05,Watling94,YANG1998} for an overview.
Ben-Akiva et al.~\cite{Ben-Akiva2002} introduced DynaMIT, a simulation-based approach designed to predict future traffic conditions.
Other works that also rely on simulation-based models include~\citep{Mahmassani2001}. 
Peeta and Mahmassani~\cite{Peeta95} introduced a rolling horizon framework addressing the real-time traffic assignment problem. 
This approach concatenates for  fixed consecutive time-intervals static flow assignments and thus does not comply to our definition of dynamic equilibrium in which at any time (also within stages) equilibrium conditions must hold.
Huang and Lam~\cite{Huang2003} allow for different user classes where each class may use a different travel time prediction.
Their model is formulated in discrete time and assumes an acyclic path formulation.

A large body of research has been dedicated to the use of deep learning techniques, in particular {\em graph neural networks} (GNNs), for predicting street segment delays in road networks.
It is impossible to list all relevant work in this section, we instead describe some key papers and point the reader to~\cite{GNNSurvey} for a complete survey.
The work in~\cite{LYSL18} uses a random walk-based graph diffusion process to create a convolutional operator that captures spatial relations.
\cite{YYZ18} propose a spatio-temporal graph convolutional network which model the temporal dependency, whereas~\cite{wu2019graph} model the spatial dependency through an adaptive learnable dependency matrix and the temporal dependency with dilated convolution~\citep{oord2016wavenet}.
Finally, {\em graph attention networks} (GATs), introduced by \cite{VCCRLB18}, have also been used in the context of traffic predictions by~\cite{ZFW020}.
We note that our work bridges the above areas of dynamic route updates based on real time information and of applying ML models for predicting traffic delay.

Gentile~\cite{Gentile2016} considered a mathematical approach incorporating traffic predictions in a DTA model.
He derives the existence of equilibria using a variational inequality approach for the considered DTA model under simplifying assumptions such as an acyclic graph.
The VI approach is arc and node-based and for its correctness, the assumption on acyclic (finite) paths is necessary as he uses a telescopic sum of edge travel times in order to arrive at a path-based VI formulation as used in~\cite{Friesz93}.
Note that this approach fails in the general setting we consider in this paper.
For further references on  adaptive route choice models we refer to \cite{KucharskiG19,Marcotte04,Hamdouch2004,UnnikrishnanW09,Watling03}.

For works analyzing the inefficiency of dynamic equilibria (within the full or instantaneous information model), we refer to  \cite{BhaskarFA15,CaoCCW17,CorreaCO19,GrafH20}.
 \section{The Flow Model}\label{sec:Model}
In the following, we describe the Vickrey fluid queuing model that we will use throughout this paper.
We consider a finite directed graph $G=(V,E)$ with positive rate capacities $\nu_e\in \R_{>0}$
and positive transit times $\tau_e\in \R_{>0}$ for every edge $e \in E$.
There is a finite set of commodities~$I$, each with a commodity-specific source node $s_i \in V$ and a commodity-specific sink node $t_i \in V\setminus\{s_i\}$.
We assume that there is at least one $s_i$-$t_i$ path for each $i \in I$ and we denote the set of nodes and edges lying on an $s_i$-$t_i$-path as $V_i$ and $E_i$, respectively.
The infinitesimal agents of every commodity $i\in I$ enter the network according to a locally integrable network inflow rate function $u_i\in\RateFcts\coloneqq \set{g\in\Lloc{1}(\R, \R_{\geq0}) \mid \restrict{g}{(-\infty, 0)} \aequal 0 }$. 

A \emph{flow over time} is a tuple $f = (f^+,f^-)$, where
$f^+, f^- \in\RateFcts^{I\times E}$ 
model the edge inflow rate $f^+_{i,e}(\theta)$ and edge outflow rate $f^-_{i,e}(\theta)$ 
of commodity $i$ of an edge $e\in E$ at time $\theta \in \R$.
The \emph{queue length} of edge $e$ at time $\theta$ is given by
\begin{align}
	q_e(\theta) \coloneqq \sum_{i \in I}F^+_{i,e}(\theta) - \sum_{i \in I}F_{i,e}^-(\theta+\tau_e) \label[cons]{eq:Cont-FlowDefProperties-QueueLengthWithF},
\end{align}
where $F_{i,e}^+(\theta):=\int_{0}^{\theta} f_{i,e}^+(z)\diff z$ and $F_{i,e}^-(\theta):=\int_{0}^{\theta}  f_{i,e}^-(z)\diff z$ 
denote the \emph{cumulative (edge) inflow} and \emph{cumulative (edge) outflow}.
For simplicity, we denote the aggregated in- and outflow rates for all commodities by $f^+_e \coloneqq \sum_{i \in I}f^+_{i,e}$ and $f^-_e \coloneqq \sum_{i \in I}f^-_{i,e}$, respectively; their cumulative variants are denoted as $F_e^+$ and $F_e^-$, respectively.

A \emph{deterministic} flow over time $f$ satisfies the following conditions \eqref{eq:Cont-FlowDefProperties-OpAtCap} and \eqref{eq:FIFO}.
We assume that the queue operates at capacity which can be modeled by requiring
\begin{align}\label[cons]{eq:Cont-FlowDefProperties-OpAtCap} 
	f_e^-(\theta) = \begin{cases}
	\capa_e, & \text{ if } q_e(\theta-\transit_e) > 0, \\
	\min\set{f^+_e(\theta-\transit_e), \capa_e}, & \text{ else, }
	\end{cases} 
\end{align}
for all  $e \in E$, $\theta \in \R$.
Moreover, we want the flow to follow a FIFO principle on the queues, which can be formalized by
	\begin{equation} \label[cons]{eq:FIFO}
		f^-_{i, e} (\theta) = 
			\begin{cases}
				f^-_e (\theta) \cdot \frac{f^+_{i, e}(\vartheta)}{f^+_e(\vartheta)}, & \text{ if }  f^+_e(\vartheta) > 0,\\
				 0, & \text{ else,}
			\end{cases}
	\end{equation}
where $\vartheta \coloneqq \min\set{\vartheta\leq \theta | \vartheta+\tau_e+\frac{q_e(\vartheta)}{\nu_e}=\theta}$ is the earliest point in time a particle can enter edge $e$ and leave at time $\theta$ while $\frac{q_e(\vartheta)}{\nu_e}$ is the current waiting time to be spent in the queue of edge $e$. 
Consequently, \cref{eq:FIFO} ensures that the share of commodity $i$ of the aggregated outflow rate at any time equals the share of commodity $i$ of the aggregated inflow rate at the time the particles entered the edge. 

A \emph{feasible flow} is a deterministic flow that also fulfills the \emph{flow conservation constraints} \eqref{eq:Cont-FlowDefProperties-FlowCons} and \eqref{eq:Cont-FlowDefProperties-FlowConsSink}.
These are modeled for commodity $i$ and node $v \neq t_i$ as
\begin{align}\label[cons]{eq:Cont-FlowDefProperties-FlowCons} 
	\sum_{e \in \delta^+_v}f^+_{i,e}(\theta) - \sum_{e \in \delta^-_v}f^-_{i,e}(\theta) = \begin{cases}
		u_i(\theta), &\text{ if }v = s_i, \\
		0,			 &\text{ if }v \neq s_i,
	\end{cases}
\end{align}
 for $\theta \in \R_{\geq 0}$ where $\delta^+_v := \set{vu \in E}$ and $\delta^-_v := \set{uv \in E}$ are the sets of outgoing edges from~$v$ and incoming edges into $v$, respectively.
 For the sink node $t_i$ of commodity~$i$ we require
	\begin{align}\label[cons]{eq:Cont-FlowDefProperties-FlowConsSink} 
		\sum_{e \in \delta^+_{t_i}}f^+_{i,e}(\theta) - \sum_{e \in \delta^-_{t_i}}f^-_{i,e}(\theta) \leq 0 & \quad \text{ for all } \theta \in \R_{\geq 0}.
	\end{align}

Cominetti et al.~\cite{CominettiCL15} analysed the queue dynamics of deterministic flows $f$:
They describe show that the queue length is given by \(
    q_e(\theta) = \max_{\xi\leq \theta}\int_{\xi}^{\theta} f_e^+ - \capa_e \diff\leb
\)
for any $\theta\in\R$, $e\in E$.
Thus, queues are independent of the edge outflow rates $f_e^-$.
Moreover, it can be shown that given a set of inflow rates $f^+ = (f_{i,e}^+)_{i, e}\in \RateFcts^{I\times E}$ there exists a unique (up to a Lebesgue-null set) family of outflow rates $f^-=(f_{i,e}^-)_{i, e}\in\RateFcts^{I\times E}$, such that $f\coloneqq (f^+, f^-)$ is deterministic; we call $f$ \emph{the deterministic flow with inflow rates $f^+$}.
If the inflow rates of any two deterministic flows $f$ and $f'$ into an edge $e$ match almost everywhere up to some time $H$, the corresponding queue length $q_e$ and exit time $T_e(\theta)\coloneqq \theta + \frac{q_e(\theta)}{\nu_e} + \transit_e$ of both flows match up to time $H$, and the outflow rates out of edge $e$ match almost everywhere up to time $T_e(H)$.

\subsection{Instantaneous Dynamic Equilibrium}

For an instantaneous dynamic equilibrium (IDE) as defined by \cite{GrafHS20} we assume that, whenever an agent arrives at an intermediate node $v\in V_i$ at time $\theta$, she is given the information about the current queue length $q_e(\theta)$ and transit time $\tau_e$ of all edges $e\in E$, and, based on this information, she computes a shortest $v$-$t_i$-path and enters the first edge on this path. We define the \emph{instantaneous travel time} of an edge~$e$ at time~$\theta$ as 
$c_e(\theta) \coloneqq \frac{q_e(\theta)}{\nu_e} + \tau_e$.
With this we can define commodity-specific node labels~$\ell_{i,v}(\theta)$ corresponding to current earliest arrival times when travelling from $v$ to the sink $t_i$ at time $\theta$ by
\begin{equation}\label{eq:current_shortest_path_distances}
    \ell_{i,v}(\theta)\coloneqq\begin{cases} \theta, & \text{ for } v=t_i,\\
    \min_{e=vw\in E_i} \ell_{i,w}(\theta)+c_{e}(\theta), & \text{ for $v\in V_i\setminus\{t_i \}$.}\end{cases}
\end{equation}
We say that edge $e=vw\in E_i$
is \emph{active} for $i\in I$ at time $\theta$,
if   $ \ell_{i,v}(\theta) = \ell_{i,w}(\theta)+c_{e}(\theta)$ and we denote the set of active edges for commodity $i$ by $E_i(\theta)\subseteq E$. 
\begin{definition}
	A feasible flow over time $f$ is an \emph{instantaneous dynamic equilibrium (IDE)}, if
	for all  $i\in I, \theta\in \R$ and $e\in E$ it satisfies 
	\begin{align}\label[cons]{eq:Cont-FlowDefProperties-OnlyUseSP}
		f_{i,e}^+(\theta)>0 \implies e\in E_i(\theta).
	\end{align}
\end{definition}

\subsection{Dynamic Nash Equilibrium}

In contrast, in the full information model we assume that agents have complete knowledge of the entire (future) evolution of the flow over time.
If an agent enters an edge $e=vw$ at time $\theta$, the travel 
time is
$c_e(\theta) \coloneqq \tau_e + \frac{q_e(\theta)}{\nu_e}$
and the  exit time of edge $e$ is given by
$ T_e(\theta) \coloneqq \theta+c_e(\theta)$. In this setting it is common (cf. \citealp{CominettiCL15}) to define the node labels in such a way as to denote the earliest possible arrival time at each node (starting from the commodity's source node). Here, however, we will instead use an equivalent definition more in line with the node labels for IDE. So, for any $i \in I, v \in V$ and $\theta \in \R$ we define a node label $\ell_{i,v}(\theta)$ denoting the earliest possible arrival time at node $t_i$ for a particle starting at time $\theta$ at node $v$ by setting
 \begin{equation}\label{t-labels-for-DE:recursive} 
     \ell_{i,v}(\theta):=\begin{cases} \theta &\text{ for } v=t_i,\\
     \min_{e=vw\in\delta^+_v} \ell_{i,w}(T_e( \theta)) &\text{ else}.
     \end{cases}
 \end{equation}
We, again, say that an edge $e=vw$ is \emph{active} for commodity $i\in I$ at time $\theta$, if it holds that $\ell_{i,v}(\theta)=\ell_{i,w}(T_e( \theta))$ and denote by $E_i(\theta)$ the set of active edges for commodity $i$ at time $\theta$.
\begin{definition}
	A feasible flow over time $f$ is a \emph{dynamic equilibrium (DE)}, if for all $e\in E, i\in I$ and $\theta\in \R$ it holds that
	\begin{equation}\label[cons]{eq:DE-t-labels} f_{i,e}^+(\theta)>0 \implies e\in E_i(\theta).\end{equation}
\end{definition}
 \section{Dynamic Prediction Equilibria}
IDE is a short-sighted behavioral concept assuming that agents at time $\bar \theta$
predict the future evolution of queue sizes according to the constant function $q_e(\theta)= q_e(\bar\theta)$ for all $\theta\geq \bar\theta$.
In the following we relax this behavioral assumption by introducing a model wherein every commodity $i\in I$ maintains a predictor $\hat q_{i,e}$ for every edge $e \in E$.
The value $\hat q_{i,e}(\theta,\bar\theta,f)$ is then the queue length at time $\theta$ on edge $e$ as predicted by commodity $i$ at time $\bar\theta$ using the historical flow over time $f$. 
Formally, a predictor $\hat q_{i,e}$ has the following signature:
    \begin{align*}
        \hat q_{i,e}:
        \R \times \R \times (\RateFcts\times \RateFcts)^{I\times E}
        \to
        \R_{\geq0}
    \end{align*}
In general, such a predictor can depend in any arbitrary way on the entire input data including, in particular, the future evolution of the flow after the prediction time $\bar\theta$.
However, for our theoretical results we require the predictors to behave in a slightly more restricted way.
The first property we require is that the predictors do not use (and, therefore, do not need) any information on the future evolution of the queues.

\newcommand{\cwaequal}{\underset{c.w.}{\aequal}}
\begin{notation}
    For a vector of functions $g=(g_i)_{i\in\firstN{d}}$ with $g_i : \R\to\R$ and some $H\in\R$, we write $\restrictUpTo{g}{H}\coloneqq(\restrict{g_i}{(-\infty, H]})_{i\in\firstN{d}}$ for the restriction of the functions to $(-\infty, H]$.

    For two vectors of functions $g=(g_i)_{i\in\firstN{d}}$, $h=(h_i)_{i\in\firstN{d}}$ with $g_i, h_i : X \to \R$, $X\subseteq \R$, we write $g \cwaequal h$ if $g_i \aequal h_i$ holds for all $i\in\firstN{d}$.
\end{notation}

\begin{definition}
    A predictor $\hat q_{i,e}$  is called \emph{causal}, if it fulfills the condition
    \[
    \restrictUpTo{f}{\bar\theta} \cwaequal \restrictUpTo{f'}{\bar\theta}
    \implies
    \predq_{i,e}(\emptyArg,\bar\theta,f)=\predq_{i,e}(\emptyArg,\bar\theta,f')
    \]
    for all $\bar\theta\in\R$ and deterministic flows $f, f'\in(\RateFcts\times\RateFcts)^{I\times E}$.
\end{definition}

The next property ensures that at any point in time there are shortest paths with respect to the predicted queue lengths that are cycle free. However, before we can formally define this property, we need some additional notation. 
If an agent of commodity $i\in I$ enters an edge $e=vw$ at time $\theta$, the predicted travel 
time estimated at time $\bar\theta$ is given by
$\hat c_{i,e}(\theta,\bar\theta, f) \coloneqq \transit_e + \frac{\hat q_{i,e}(\theta,\bar\theta, f)}{\capa_e}$
and the predicted exit time of edge $e$ is given by
$ \hat T_{i,e}(\theta,\bar\theta, f) \coloneqq \theta+\hat c_{i,e}(\theta,\bar\theta, f)$.
We call these times \emph{$\bar\theta$-estimated} to emphasize that these values are predictions
made at time~$\bar\theta$.
\begin{definition}
A predictor  $\hat q_{i,e}$ \emph{respects FIFO} if for any edge $e$, deterministic flow $f$ and prediction time $\bar{\theta}$ the predicted exit time $\hat T_{i,e}(\emptyArg,\bar{\theta},f)$ is non-decreasing.
\end{definition}
This now allows us to describe how agents determine routes according to the predicted queues.
At time $\bar\theta$ an agent of commodity $i\in I$ predicts that if she enters a path $P=(e_1,\dots,e_k)$ at time $\theta$ she will arrive at the endpoint of $P$ at time
\begin{equation} 
    \hat \ell_i^P(\theta,\bar\theta, f) \coloneqq
    \left(\hat T_{i,e_k}(\emptyArg, \bar\theta, f)\circ\cdots\circ \hat T_{i,e_1}(\emptyArg, \bar\theta, f)\right)(\theta).
\end{equation}

 Denoting the (finite) set of all simple $v$-$t_i$ paths by $\paths_{i,v}$, the earliest $\bar\theta$-estimated time at which an agent starting at time $\theta$
 from node $v\in V_i$ can reach $t_i$ is given by
 \begin{equation}\label{path:labels-ML} 
 \hat\ell_{i,v}(\theta,\bar\theta, f):=\min_{P\in \paths_{i,v}}\hat \ell^P_i(\theta,\bar\theta, f).
 \end{equation}
If all predictors respect FIFO, the label functions defined in~\eqref{path:labels-ML} satisfy the equations
 \begin{equation}\label{labels-ML:recursive}
 \hat\ell_{i,v}(\theta, \bar\theta, f)=\begin{cases} \theta, &\text{ if } v=t_i,\\
 \displaystyle\min_{vw\in\delta^+_v} \hat\ell_{i,w}( \hat T_{i,vw}( \theta,\bar\theta, f),\bar\theta, f ), &\text{ if } v\neq t_i.
 \end{cases}
 \end{equation}
Moreover, the \emph{predicted delay for taking an edge $e=vw\in E_i$ at time $\theta$} is then defined as
\[ 
    \predDel_{i,e}(\theta,\bar\theta,f) \coloneqq \predl_{i,w}\left(\predT_{i,e}(\theta,\bar\theta,f), \bar\theta, f \right) - \predl_{i,v}(\theta,\bar\theta,f).
\]
 We say that an edge $e=vw\in E_i$ is \emph{$\bar\theta$-estimated active} for commodity $i$ at time $\theta$, if $\predDel_{i,e}(\theta,\bar\theta,f) = 0$ holds true.
Furthermore, let us denote the set of $\bar\theta$-estimated active edges for commodity $i$ at time $\theta$ by $\hat E_i(\theta,\bar\theta, f)$.

\begin{definition}
A pair $(\hat q, f)$ of a set of predictors $\hat q = (\hat q_{i,e})_{i\in I, e\in E}$ and a feasible flow over time $f$ is a \emph{dynamic prediction equilibrium (DPE)}, if for all $e\in E, i\in I$ and $\theta\geq 0$ it holds 
\begin{equation}\label{dpe-property} f_{i,e}^+(\theta)>0  \implies  e\in \hat E_i(\theta,\theta, f).\end{equation}
We then call the flow $f$ a \emph{dynamic prediction flow with respect to the predictor $\hat q$}. 
\end{definition}

In other words, in a DPE, particles only follow predicted shortest paths, and particles predict a shortest path (according to their commodity's predictor) when they enter the network and every time they arrive at an intermediate node $v\neq t_i$. 
 \section{Existence of Dynamic Prediction Equilibria}

In this section we show that for causal predictors that respect FIFO and a suitable continuity property there always exists a dynamic prediction equilibrium. We will also give several examples of such predictors, including one for which any DPE corresponds to an IDE and vice versa.

\subsection{Existence of DPE Using a Variational Inequality}\label{sec:existence-using-variational-inequality}

To show the existence of DPE we make use of a result by \citet[Theorem 24]{brezis1968} guaranteeing the existence of solutions to certain variational inequalities on a reflexive Banach space $X$.
For this, we call a mapping $\mathcal A: K\to Y$ from $K\subseteq X$ to the normed space $Y$ \emph{sequentially weak-strong continuous} if $\mathcal A$ maps every sequence $(u_i)_i$ in $X$ that \emph{converges weakly to $u\in X$}, i.e., $\lim_{i\to\infty}\scalprod{g}{u_i} = \scalprod{g}{u}$ for all $g\in X'$, to a strongly convergent sequence, i.e., $\lim_{i\to\infty}\norm{A(u_i) - A(u)}_Y=0$.
Here, $\scalprod{\emptyArg}{\emptyArg}$ denotes the canonical pairing between $X$ and its continuous dual space $X'$.
\citet[Section~5]{CominettiCL15} have found the following useful special case of Brézis's theorem.

\begin{theorem}\label{thm:var-ineq}
    Let $X$ be a reflexive Banach space and let $\mathcal{A}: K\to X'$ be a sequentially weak-strong continuous map defined on a non-empty, closed, bounded and convex set $K\subseteq X$.
    Then there exists a solution $u\in X$ to the variational inequality \begin{equation}\label{eq:var}
        \forall v \in X: \quad
        \scalprod{\mathcal A(u)}{v - u} \geq 0
        .\end{equation}
\end{theorem}

Here, we apply this theorem on the reflexive Banach space $X= L^p(S)^d$ with $S$ a compact set in $\R^2$ and $1<p<\infty$.
Its continuous dual space is isomorphic to $X' = L^q(S)^d$, where $q$ is the conjugate of $p$ with $\nicefrac{1}{p} + \nicefrac{1}{q} = 1$.
The canonical pairing is then given by $\scalprod{g}{f}\coloneqq \sum_{i\in\firstN{d}} \int_S g_i\cdot f_i\diff\leb$ for $f\in X$ and $g\in X'$.

This theorem can be used to build up a dynamic prediction flow with respect to a given set of predictors by iteratively extending so-called \emph{partial dynamic prediction flows} which fulfill the equilibrium property up to some time horizon.
A sufficient regularity condition on the predictors to apply the theorem is depicted in the following.
\begin{definition}
    A predictor $\hat q_{i,e}$ is \emph{$p$-continuous} for $1 < p < \infty$, if  $\predq_{i,e}(\emptyArg,\emptyArg,f)$ is continuous for every deterministic flow $f$ with $f^+\in \Lloc{p}(\R,\R_{\geq 0})^{I\times E}$ and
    $$L^p([0, M], \R_{\geq0})^{I\times E} \to C([0,M]\times D, \R_{\geq0}), \quad f^+\mapsto \predq_{i,e}(\emptyArg,\emptyArg, f)$$
    is sequentially weak-strong continuous for every $M > 0$ and compact interval $D$.
    Here, $f$ denotes the deterministic flow with inflow rates $\CharF{[0,M]}\cdot f^+$.
\end{definition}

Now, we formally introduce the concept of partial dynamic prediction equilibria.
\begin{definition}
    A deterministic flow $f$ is \emph{feasible up to time $H$} if $f$ fulfills the flow conservation conditions~\eqref{eq:Cont-FlowDefProperties-FlowCons} and \eqref{eq:Cont-FlowDefProperties-FlowConsSink} for $\theta\leq H$.
    We call $f$ a \emph{partial dynamic prediction flow with respect to a set of predictors $\hat q$ up to time $H$} if $f$ is feasible up to time $H$ and $f_{i,e}^+(\theta)>0$ implies $e\in \hat E_i(\theta,\theta,f)$ for all $\theta \leq H, e \in E, i \in I$.
\end{definition}

We now show that the time horizon $H$ of any partial dynamic prediction flow can be extended.
We employ a proof-technique similar to the one used in \cite[Lemma~5.6]{GrafHS20} for the extension property of IDE flows.
However, the analysis is more involved as we allow for a more general functional dependence of the predicted queue lengths on the past flow evolution.
This stands in contrast to IDEs where each prediction only depends on the queue lengths of one edge at a single point in time.

\begin{lemma}\label{lem:extension}
    Let $u_i$ be network inflow rates in $\RateFcts\cap \Lloc{p}(\R,\R_{\geq0})$ and $\hat q = (\hat q_{i,e})_{i,e}$ be a set of $p$-continuous and causal predictors that respect FIFO.
    We can extend any partial dynamic prediction flow $f$ with respect to $\hat q$ up to time $H$ to a dynamic prediction flow up to time $H + \mintransit$ with $\mintransit\coloneqq \min_{e\in E} \transit_e$.
\end{lemma}

The main idea of the proof is to first define a set $K$ of all possible extensions of the given partial flow.
We then define a mapping $\mathcal{A}: K \to L^q(D)^{I\times E}$ associating with each possible extension $g$ a function which for every commodity $i$, edge $e$ and time $\theta$ returns the predicted delay $\predDel_{i,e}(\theta,\theta, g)$.
The $p$-continuity of the predictors implies that $\mathcal A$ is sequentially weak-strong continuous such that \Cref{thm:var-ineq} gives a solution to the variational inequality~\eqref{eq:var}.
This solution is an extension of $f$ which satisfies the properties of a partial dynamic prediction flow up to time $H+\mintransit$.

In this section, we equip the space of continuous functions $C(S)$ on a compact set $S\subseteq \R^d$ with the topology induced by the uniform norm $\norm{g}_\infty\coloneqq \max_{x\in S} g(x)$.

\begin{lemma}\label{lem:delay-weak-strong-cont}
    If the predictors $(\predq_{i,e})_{e\in E}$ are $p$-continuous, the mappings
    \begin{enumerate}[label=(\roman*)]
        \item\label{T_e-weak-strong} $f^+\mapsto \predT_{i,e}(\emptyArg,\emptyArg, f)$ for all $e\in E$,
        \item\label{l^P-weak-strong} $f^+\mapsto \predTP iP(\emptyArg,\emptyArg, f)$ for all simple paths $P$,
        \item\label{l_v-weak-strong} $f^+\mapsto \predl_{i,v}(\emptyArg,\emptyArg, f)$ for all nodes $v\in V_i$, and
        \item\label{delta-weak-strong} $f^+\mapsto \predDel_{i,e}(\emptyArg, \emptyArg, f)$ for all $e\in E_i$.
    \end{enumerate}
    are sequentially weak-strong continuous from $L^p([0, M],\R_{\geq0})^{I\times E}$ to $C([0,M]\times D)$ for $M > 0$ and compact intervals $D$.
    Here, $f$ denotes the deterministic flow with inflow rates $\CharF{[0,M]}\cdot f^+$.
\end{lemma}
\begin{proof}
    \ref{T_e-weak-strong}.
    The sequential weak-strong continuity of $f^+\mapsto \predT_{i,e}(\emptyArg,\emptyArg, f)$ directly follows from the $p$-continuity of $\predq_{i,e}$.

    \ref{l^P-weak-strong}.
    For $f^+\mapsto \predTP iP(\emptyArg,\emptyArg, f)$ we use induction on the length of the simple path $P$.
    In the base case, $P$ is an empty path such that $\predTP iP(\emptyArg, \emptyArg, f) = ((\theta,\bar\theta)\mapsto \theta)$ for any deterministic flow $f$, and $f^+\mapsto \predTP iP(\emptyArg,\emptyArg, f)$ is a constant (and thus continuous) map.
    Assume that $P$ consists of some path $P'$ and an additional final edge $e$ and let $(f^{+,k})_{k\in\N}$ be weakly converging to some $f^+$.
    We want to show that \begin{align*}
             & \norm{ \predTP iP(\emptyArg,\emptyArg, f^k) - \predTP iP(\emptyArg,\emptyArg, f) }_\infty \\
             & = \max_{\substack{\theta\in[0,M]                                                          \\\bar\theta\in D}} \abs{
                \predT_{i,e}\left( \predTP i{P'}(\theta,\bar\theta,f^k), \bar\theta, f^k \right)
                -
                \predT_{i,e}\left( \predTP i{P'}(\theta,\bar\theta,f), \bar\theta, f \right)
            }
        \end{align*}
        converges to $0$ for $k\to\infty$.
        We split up this term using the triangle inequality to get
        \begin{align*}
             & \max_{\substack{\theta\in[0,M]                            \\\bar\theta\in D}} \abs{
                \predT_{i,e}\left(\predTP i{P'}(\theta,\bar\theta, f^k), \bar\theta, f^k\right)
                -
                \predT_{i,e}\left(\predTP i{P'}(\theta,\bar\theta, f), \bar\theta, f\right)
            }                                                            \\
             & \leq \overbrace{\max_{\substack{\theta\in[0,M]            \\\bar\theta\in D}} \abs{
                    \predT_{i,e}\left(\predTP i{P'}(\theta,\bar\theta, f^k), \bar\theta, f^k\right)
                    -
                    \predT_{i,e}\left(\predTP i{P'}(\theta,\bar\theta, f^k), \bar\theta, f\right)
            }}^{\eqqcolon \alpha^k}                                      \\
             & \phantom{==} + \underbrace{\max_{\substack{\theta\in[0,M] \\\bar\theta\in D}} \abs{
                    \predT_{i,e}\left(\predTP i{P'}(\theta,\bar\theta, f^k), \bar\theta, f\right)
                    -
                    \predT_{i,e}\left(\predTP i{P'}(\theta,\bar\theta, f), \bar\theta, f\right)
                }}_{\eqqcolon \beta^k}.
        \end{align*}
    By induction there exists an $N\in\N$ such that \displayStyleForArxiv{\ensuremath{
            \smallnorm{\predTP i{P'}(\emptyArg,\emptyArg, f^k) - \predTP i{P'}(\emptyArg,\emptyArg, f)}_\infty < 1
        }} holds for all $k\geq N$.
    With $M' \coloneqq \smallnorm{\predTP i{P'}(\emptyArg,\emptyArg, f)}_{[0,M]\times D} + 1$, the value $\predTP i{P'}(\theta,\bar\theta, f^k)$ is contained in $[0, M']$ for all $k\geq N$, $\theta\in[0,M]$, and $\bar\theta\in D$.
    Moreover, $\CharF{[0,M]}\cdot f^{+,k}$ converges weakly to $\CharF{[0,M]}\cdot f^+$ in $L^p([0, M'])^{I\times E}$.
    Using \ref{T_e-weak-strong} for $M'$ we conclude
    \[
        \alpha^k \leq \norm{ \predT_{i,e}(\emptyArg,\emptyArg, f^k) - \predT_{i,e}(\emptyArg,\emptyArg, f) }_{C([0, M']\times D)} \xrightarrow[k\to\infty]{} 0
    .\]
    The term $\beta_k$ vanishes because of the uniform continuity of $\predT_{i,e}(\emptyArg,\emptyArg,f)$ on the compact set $[0,M']\times D$.

    \ref{l_v-weak-strong}.
    Let $(f^{+,k})_{k\in\N}$ weakly converge to $f^+$ in $L^p([0,M], \R_{\geq0})^{I\times E}$.
    By \ref{l^P-weak-strong}, the sequence $(\predTP iP(\emptyArg,\emptyArg, f^k))_k$ uniformly converges to $\predTP iP(\emptyArg,\emptyArg,f)$ on $C([0,M]\times D,\R_{\geq0})$ for any simple $v$-$t_i$-path $P$.
    The minimum of these functions also converges uniformly.

    \ref{delta-weak-strong}.
    Let $(f^{+,k})_{k\in\N}$ weakly converge to $f^+$ in $L^p([0,M], \R_{\geq0})^{I\times E}$.
    For $e=vw\in E_i$, we show \displayStyleForArxiv{\ensuremath{
            \smallnorm{\predDel_{i,e}(\emptyArg,\emptyArg, f^k) - \predDel_{i,e}(\emptyArg,\emptyArg, f)}_\infty \rightarrow 0.
        }}
    In a first step, we split up the sum again using the triangle inequality to obtain
    \begin{align*}
         & \norm{\predDel_{i,e}(\emptyArg,\emptyArg, f^k) - \predDel_{i,e}(\emptyArg,\emptyArg, f)}_\infty    \\
         & \leq \norm{ \predl_{i,v}(\emptyArg,\emptyArg, f^k) - \predl_{i,v}(\emptyArg,\emptyArg, f) }_\infty \\
         & \phantom{==}+\max_{\substack{\theta\in[0,M]                                                        \\\bar\theta\in D}} \abs{
            \predl_{i,w}\left(\predT_{i,e}(\theta,\bar\theta,f^k), \bar\theta, f^k\right)
            -
            \predl_{i,w}\left(\predT_{i,e}(\theta,\bar\theta,f), \bar\theta, f\right).
        }
    \end{align*}
    The first term converges to $0$ by \ref{l_v-weak-strong}.
    The second term is handled as in~\ref{l^P-weak-strong}:
        We apply the triangle inequality once again to get
        \begin{align*}
             & \max_{\substack{\theta\in[0,M]                            \\\bar\theta\in D}} \abs{
                \predl_{i,w}\left(\predT_{i,e}(\theta,\bar\theta,f^k), \bar\theta, f^k\right)
                -
                \predl_{i,w}\left(\predT_{i,e}(\theta,\bar\theta,f), \bar\theta, f\right)
            }                                                            \\
             & \leq \overbrace{\max_{\substack{\theta\in[0,M]            \\\bar\theta\in D}} \abs{
                    \predl_{i,w}\left(\predT_{i,e}(\theta,\bar\theta,f^k), \bar\theta, f^k\right)
                    -
                    \predl_{i,w}\left(\predT_{i,e}(\theta,\bar\theta,f^k), \bar\theta, f\right)
            }}^{\eqqcolon\alpha^k}                                       \\
             & \phantom{==} + \underbrace{\max_{\substack{\theta\in[0,M] \\\bar\theta\in D}} \abs{
                    \predl_{i,w}\left(\predT_{i,e}(\theta,\bar\theta,f^k), \bar\theta, f\right)
                    -
                    \predl_{i,w}\left(\predT_{i,e}(\theta,\bar\theta,f), \bar\theta, f\right)
                }}_{\eqqcolon\beta^k}.
        \end{align*}
        By \ref{T_e-weak-strong} there exists an $N\in\N$ with $\smallnorm{\predT_{i,e}(\emptyArg,\emptyArg, f^k) - \predT_{i,e}(\emptyArg,\emptyArg, f)}_{[0,M]\times D} < 1$ for all $k\geq N$
        and define $M' = \smallnorm{\predT_{i,e}(\emptyArg,\emptyArg,f)}_{[0,M]\times D} + 1$.
        Then $\predT_{i,e}(\theta,\bar\theta, f^k)\in [0,M']$ for any $\theta\in[0,M],\bar\theta\in D$ for all $k\geq N$.
        Using \ref{l_v-weak-strong}, we conclude
        \[
            \alpha^k \leq \norm{ \predl_{i,w}(\emptyArg,\emptyArg, f^k) - \predl_{i,w}(\emptyArg,\emptyArg, f) }_{C([0, M']\times D)} \xrightarrow[k\to\infty]{} 0
            .\]
        The term $\beta_k$ converges to $0$ because of the uniform continuity of $\predl_{i,w}(\emptyArg,\emptyArg,f)$ on the compact set $[0,M']\times D$.
\end{proof}

With this \namecref{lem:delay-weak-strong-cont} we can now turn back to the proof of the extension-lemma:

\begin{proof}[Proof of \Cref{lem:extension}]
    \newcommand{\badTimes}{\Phi}
    Let $f$ be a partial dynamic prediction flow up to time $H$ with respect to $\predq$.
    The outflow rate on an edge $e$ of any deterministic flow $h$ whose inflow rates coincide with $f$ up to time $H$ are already uniquely determined up to time $T_e(H) \geq H + \tau_e$.
    Therefore, the rates $b_{i,v}^-(\theta) \coloneqq \sum_{e\in\inEdges{v}} \outfl[h]_{i,e}(\theta) + \one_{v=s_i} u_i(\theta)$ on the interval $D\coloneqq [H, H+\mintransit]$ are independent of the edge inflow rates $\restrict{\infl[h]_{i,e}}{D}$ on $D$.

    We now want to apply Brézis's theorem in the form of \Cref{thm:var-ineq} to find suitable inflow rates $\restrict{\infl[h]_{i,e}}{D}$.
    For that, we define the set $K\subseteq L^p(D)^{I\times E}$ as follows:
    \[
        K\coloneqq \Set{ g\in L^p(D, \R_{\geq 0})^{I\times E} | \begin{array}{c r c l}
                \forall i\in I, v\in V\setminus\{t_i\} : & \sum_{e\in\outEdges{v}} g_{i,e}   & \aequal & b_{i,v}^-,  \\
                \forall i\in I :                         & \sum_{e\in\outEdges{t_i}} g_{i,e} & \aleq   & b_{i,t_i}^- \\
            \end{array}
        }.
    \]

    The elements of $K$ are the possible inflow rates for the interval $D$ that fulfill the flow conservation constraints.
    More specifically, for any $g\in K$, let $\bar g$ denote the unique deterministic flow with inflow rates \[
        \infl[\bar g]_{i,e}(\theta) \coloneqq \begin{cases}
            g_{i,e}(\theta),     & \text{if $\theta\in D$,} \\
            \infl_{i,e}(\theta), & \text{otherwise.}
        \end{cases}
    \]

    \begin{claim}\label{claim:feasible-and-dpe-up-to-h}
        For each $g\in K$, $\bar g$ is a partial dynamic prediction flow up to time $H$ and feasible up to time $H+\mintransit$.
    \end{claim}
    \begin{proofClaim}
        As $f$ is feasible up to time $H$ and $g\in K$, it follows that $\bar g$ is feasible up to time $H+\mintransit$.
        Because all predictors are causal, the equilibrium property \[
            \infl[\bar g]_{i,e}(\theta) > 0 \,\implies\, e \in \predE_i(\theta, \theta, \bar{g})
        \]
        is transferred from $f$ over to $g$ for $\theta < H$.
    \end{proofClaim}

    \begin{claim}
        The set $K$ is nonempty, closed, bounded and convex.
    \end{claim}
    \begin{proofClaim}
        By $u_i\in \Lloc{p}(\R, \R_{\geq0})$ and $\outfl_{i,e}(\theta)\leq \capa_e$, it follows that $b^-_{i,v}$ is $p$-integrable on $D$ for all $v\in V$.
        Then, $g\in K$ implies $\smallnorm{g_{i,e}}_{L^p(D)} \leq \smallnorm{b^-_{i,v}}_{L^p(D)}$ for all $i\in I$ and $e=vw\in E$.
        This implies \begin{align*}
                \norm{g}_{L^p(D)^{I\times E}}
                 & = \sum_{i, e} \norm{g_{i,e}}_{L^p(D)}
                = \sum_{i, v} \sum_{e\in \outEdges{v}} \norm{g_{i,e}}_{L^p(D)}   \\
                 & = \sum_{i,v} \norm{\sum_{e\in\outEdges{v}} g_{i,e} }_{L^p(D)}
                \leq \sum_{i,v} \norm{b^-_{i,v}}_{L^p(D)}
                < \infty.
            \end{align*}
        Therefore, $K$ is bounded in $L^p(D)^{I\times E}$.
        To see that $K$ is nonempty, observe that, in a partial dynamic prediction flow, particles of commodity $i$ can only arrive at a node $v$ if $t_i$ is reachable from $v$.
        Hence, for all $v\in V\setminus\{ t_i\}$ with $b_{i,v}^-(\theta)> 0$ we select an arbitrary edge $e\in\outEdges{v}$ and set $g_{i,e} = b_{i,v}^-$.
        For all other edges, we set $g_{i,e} = 0$.
        This implies $g\in K$.
        It can be easily verified that $K$ is convex and closed.
    \end{proofClaim}

    By \Cref{claim:feasible-and-dpe-up-to-h}, $\bar g$ is a dynamic prediction flow up to time $H$ for every $g\in K$.
    Therefore, we are looking for some $g\in K$ such that \[
        \infl[\bar g]_{i,e}(\theta) > 0 \,\implies\, e \in \predE_i(\theta, \theta, \bar{g})
    \]
    is also fulfilled for $\theta\in D$.
    For an edge $e=vw\in E_i$ lying on a directed $s_i$-$t_i$-path, the statement $e \in \predE_i(\theta,\theta,\bar g)$ is equivalent to $\predDel_{i,e}(\theta,\theta,\bar g)\leq 0$.
    Using this observation, we define $\mathcal A: K \to L^q(D)^{I\times E}$ with $\nicefrac{1}{p} + \nicefrac{1}{q} = 1$ as the predicted delay operator when using an edge $e=vw$:
        \[
            \mathcal{A}(g)_{i,e}(\theta) \coloneqq \begin{cases}
                \predDel_{i,e}(\theta,\theta, \bar g), & \text{if $e\in E_i$,} \\
                1,                                     & \text{otherwise.}
            \end{cases}
        \]
        We note that for edges $e\notin E_i$ that are irrelevant to a commodity $i$, we simply set the predicted delay to $1$;
        however, any strictly positive constant would work.
    By the $p$-continuity of $\predq$, the function $\mathcal A(g)_{i,e}$ is continuous as a function on $\R$.
    Thus, $\mathcal A(g)$ is indeed contained in $L^q(D)^{I\times E}$ as a function on $D$.

    \begin{claim}
        The map $\mathcal A$ is non-negative and sequentially weak-strong continuous.
    \end{claim}
    \begin{proofClaim}
        The non-negativity of $\mathcal A$ results from the predictors respecting FIFO.

        To show that $\mathcal{A}$ is sequentially weak-strong continuous, let $(g^k)_{k\in\N}$ be a sequence in $L^p(D)^{I\times E}$ that converges weakly to $g^+$.
        Then, $\restrict{\bar g^{k, +}}{[0, M]}$ also converges weakly to
        $\restrict{\bar g^+}{[0, M]}$ in $L^p([0, M])^{I\times E}$ with $M\coloneqq H+\mintransit$.
        By \Cref{lem:delay-weak-strong-cont} and the causality of the predictors, the sequence $\predDel_{i,e}(\emptyArg,\emptyArg,\bar g^k)$ converges to $\predDel_{i,e}(\emptyArg,\emptyArg, \bar g)$ with respect to the uniform norm on $C([0,M]\times D)$ for any $e\in E_i$.
        We use this information to compute the convergence on the space $C(D)^{I\times E}$:
        \begin{align*}
            \norm{\mathcal A(g^k) - \mathcal A(g) }_\infty
                & = \max_{\substack{i\in I    \\ e\in E_i}} \max_{\theta\in D} \abs{
                \predDel_{i,e}(\theta,\theta,\bar g^k) - \predDel_{i,e}(\theta,\theta, \bar g)
            }                              \\
                & \leq \max_{\substack{i\in I \\ e\in E_i}} \norm{
                \predDel_{i,e}(\emptyArg,\emptyArg, \bar g^k) - \predDel_{i,e}(\emptyArg,\emptyArg,\bar g)
            }_{C([0,M]\times D)} \xrightarrow[k\to\infty]{} 0.
        \end{align*}
        This also implies convergence on $L^q(D)^{I\times E}$:
        For arbitrary $\varepsilon$ there exists $N\in\N$ with $\smallnorm{\mathcal A(g^k) - \mathcal A(g)}_\infty < \varepsilon/\mintransit$ for all $k\geq N$.
        This implies \[
            \norm{\mathcal A_{i,e}(g^k) - \mathcal A_{i,e}(g)}_{L^q(D)} \leq \mintransit\cdot \norm{\mathcal A_{i,e}(g^k) - \mathcal A_{i,e}(g)}_\infty < \varepsilon.
        \]
    \end{proofClaim}

    The variational inequality as given in \Cref{thm:var-ineq} now admits a solution $g\in K$ such that for all $h\in K$ we have \(
    \scalprod{\mathcal{A}(g)}{h - g} \geq 0
    .\)
We assume that the equilibrium property for $\bar g$ does not hold almost everywhere on $D$, implying that the union of all sets\displayStyleForArxiv{\ensuremath{
        \badTimes_{i,e} \coloneqq
        \Set{
            \theta \in D
            |
            \infl[\bar g]_{i,e}(\theta) > 0,\mathcal{A}(g)_{i,e}(\theta)> 0
        }
    }}
    has positive measure.
    We now construct flow rates $h\in K$ that lead to the contradiction $\scalprod{\mathcal{A}(g)}{h-g} < 0$.
    To do this, we observe that by the continuity of the maps $(\theta,\bar\theta) \mapsto \predl_{i,w}(\theta,\bar\theta, \bar g)$ and $\theta \mapsto \predT_{i,e}(\theta,\theta,\bar g)$ the set of times at which an edge $e$ is active, denoted by
    \[
        \Theta_{i,e} \coloneqq \Set{
            \theta\in \R
            |
            e \in \predE_{i,e}(\theta,\theta, \bar g)
        },
    \]
    is closed and thus measurable.
        We now define \[
            h_{i,e} :D \to \R, \quad\theta \mapsto \frac{b_{i,v}^-(\theta)}{\abs{\outEdges{v}\cap\predE_{i}(\theta,\theta, \bar g)}} \begin{cases}
                \frac{b_{i,v}^-(\theta)}{\abs{\outEdges{v}\cap\predE_{i}(\theta,\theta, \bar g)}}, & \text{if $\theta \in \Theta_{i,e}$,} \\
                0,                                                                                 & \text{otherwise,}
            \end{cases}
        \]
        for all $i\in I$, $e=vw\in E$.
        This function is measurable, because the function $\theta\mapsto \smallabs{\outEdges{v}\cap \predE_i(\theta,\theta,\bar g)}$ can be expressed as $\sum_{e\in\outEdges{v}} \CharF{\Theta_{i,e}} $, and we have $h\in L^p(D)^{I\times E}$.
        As the set $\outEdges{v}\cap\predE_{i}(\theta,\theta, \bar g)$ is non-empty for any $\theta\in D$ and $v\in V_i\setminus\{t\}$, it follows that \[
            \sum_{e\in\outEdges{v}} h_{i,e}(\theta)
            = \sum_{e\in\outEdges{v}\cap \predE_i(\theta,\theta, \bar g)} \frac{b_{i,v}^-(\theta)}{\abs{\outEdges{v}\cap \predE_i(\theta,\theta, \bar g)}} \,\begin{cases}
                = b_{i,v}^-(\theta),    & \text{if $v\in V\setminus\{ t \}$}, \\
                \leq b_{i,v}^-(\theta), & \text{otherwise,}
            \end{cases}
        \]
        because $b_{i,v}^-(\theta)$ can only be positive for nodes in $V_i$.
        Therefore, $h$ is an element of $K$.
    
    By definition, $h_{i,e}(\theta)$ is only positive if $e$ is active at time $\theta$ (w.r.t. $\bar g$) which implies $\mathcal A(g)_{i,e}(\theta) = 0$.
    Therefore, we conclude
    \begin{align*}
        \scalprod{\mathcal{A}(g)}{h-g}
        = \sum_{i,e} \int_D \mathcal{A}(g)_{i,e} \cdot h_{i,e} \diff\leb - \scalprod{\mathcal A(g)}{g}
        = - \sum_{i,e}\int_D \mathcal{A}(g)_{i,e} \cdot g_{i,e} \diff\leb.
    \end{align*}
    
    Let $i\in I$ and $e\in E$ such that $\badTimes_{i,e}$ has positive measure.
    Because both $g_{i,e}$ and $\mathcal A(g)_{i,e}$ are positive on $\badTimes_{i,e}$, this implies $\int_{\badTimes_{i,e}} \mathcal A(g)_{i,e} \cdot g_{i,e} \diff\leb > 0$.
    The non-negativity of $\mathcal A(g)$ and $g$ now implies\displayStyleForArxiv{\ensuremath{
            \scalprod{\mathcal A(g)}{h-g} \leq -\int_{\badTimes_{i,e}} \mathcal A(g)_{i,e}\cdot g_{i,e}\diff\leb < 0,
        }}
        which contradicts that $g$ is a solution to the variational inequality.
\end{proof}

With this, we can show the existence of dynamic prediction flows for causal and $p$-continuous predictors respecting FIFO.

Starting with the zero-flow with horizon $0$ and iteratively extending it using \Cref{lem:extension}, we obtain a (complete) dynamic prediction flow, as shown in the following theorem. Note, that this result allows different commodities to use different sets of predictors.

\begin{theorem}\label{thm:existence}
    For any network with a finite set of commodities, each associated with a network inflow rate in $\Lloc{p}(\R,\R_{\geq0})$ and causal and $p$-continuous predictors $\hat q_{i,e}$ that respect FIFO, there exists a dynamic prediction flow with respect to $\hat q$.
\end{theorem}
\begin{proof}
    Let $f^0\in (\RateFcts\times \RateFcts)^{I\times E}$ with $f^{0,+}_{i,e}, f^{0,-}_{i,e} \coloneqq 0$ for all $i\in I$, $e\in E$.
    Then $f^0$ is a partial dynamic prediction flow up to time $0$.
    Recursively define $f^{n+1}$ as the dynamic prediction flow up to time $\mintransit\cdot(n+1)$ given through the extension of $f^n$ as per \Cref{lem:extension}.
    It follows that $\restrictUpTo{f^{n}}{k\cdot \mintransit} \cwaequal \restrictUpTo{f^{k}}{k\cdot\mintransit}$ whenever $n\geq k$.

    We define the flow $f^\infty$ using \[
        f^{\infty, +}_{i,e}(\theta) \coloneqq f_{i,e}^{n_\theta,+}(\theta)
        \quad\text{and}\quad
        f^{\infty, -}_{i,e}(\theta) \coloneqq f_{i,e}^{n_\theta,-}(\theta)
        \quad
        \text{with $\textstyle n_\theta\coloneqq\ceil{\frac{\theta}{\mintransit}}$}
    \]
    for all $\theta \geq 0$.
    Then $f^\infty$ is a dynamic prediction flow with respect to $\predq$ up to time $\infty$:
    The feasibility and the equilibrium properties are checked by using $\restrictUpTo{f^{\infty}}{n\cdot \mintransit} \cwaequal \restrictUpTo{f^{n}}{n\cdot\mintransit}$, the fact that $f^{n}$ fulfills the desired properties for all $n\in\N$, and the causalilty of the predictors.
\end{proof}

\begin{example}
        To see why we require the predictors to be continuous, consider the non-continuous predictor        
        \[\hat q_e(\theta,\bar{\theta},f) \coloneqq \begin{cases}
                    q^f_e(\bar\theta), & \text{ if } q^f_e(\bar\theta) < 1 \\
                    2,                 & \text{ otherwise.}
                \end{cases}\]
        
        Using this predictor in a network consisting of only a single source-sink pair connected by two parallel edges $e_1$ and $e_2$ (see \Cref{fig:counter-example-continuous-predictor})
        can already lead to a situation where no equilibrium flow exists. Let $\nu_{e_1} = 1, \tau_{e_1} = 1, \nu_{e_2} = 2, \tau_{e_2} = 2$ and assume a constant inflow rate of $2$ at the source. Assume a dynamic prediction flow $f$ exists.
        Then, clearly, during the time interval $[0,1)$ agents using the above predictor may only enter edge $e_1$ (as the predicted travel time along edge $e_1$ is strictly smaller than $2$), and thus $q^f_{e_1}(1)=1$.
        Let $O\subseteq [1,\infty]$ denote the open set containing all times $\theta\geq 1$ with $q_{e_1}^f(\theta) < 1$, and let $O=\bigcup_{j\in J} (a_j, b_j)$ be a partition of $O$ into at most countably many disjoint, non-empty, open intervals.
        For every $j\in J$, we have $a_j\not\in O$, and thus $q^f_{e_1}(a_j) = 1$ holds by the continuity of $q_{e_1}^f$.
        During $(a_j, b_j)$, $e_1$ is the only active edge, and thus, for any $\theta\in (a_j,b_j)$, it follows $1>q_{e_1}^f(\theta)=q_{e_1}^f(a_j) + (\theta - a_j) >1$. Thus, $O$ is the empty set.
        This means, starting from time $1$ all particles would choose edge $e_2$, implying $q_{e_1}^f(\theta) < 1$ for all $\theta > 1$; a contradiction.
        \begin{figure}
            \centering
            \begin{tikzpicture}
                \node[draw,circle](s)at(0,0) {$s$};
                \node[draw,circle](t)at(3,0) {$t$};

                \draw[thick,->](s) to[bend left=30]node[above]{$(1,1)$} (t);
                \draw[thick,->](s) to[bend right=30]node[below]{$(2,2)$} (t);
            \end{tikzpicture}
            \caption{A network where the use of a non-continuous predictor can result in a situation where no dynamic prediction equilibrium exists. Edges are labeled with $(\tau_e, \nu_e)$.}\label{fig:counter-example-continuous-predictor}
        \end{figure}
\end{example}

\subsection{A Sufficient Regularity Condition}

To make the previous theorem more accessible, we give a sufficient, less abstract regularity condition for  $p$-continuity.

\begin{definition}\label{def:sufficient-regularity-condition}
    A predictor $\predq_{i,e}$ \emph{depends continuously on the cumulative inflow, total cumulative outflow and queue lengths},
    if there exists a continuous map \[
        \gamma_{i,e}: \R\times\R\times C(\R, \R_{\geq 0})^{(I\times E) + E + E} \to \R_{\geq0}
    \]
    such that $\predq_{i,e}(\theta,\bar\theta, f)=
        \gamma_{i,e}(
        \theta,
        \bar\theta,
        F_{I\times E}^{+,f},
        F_{E}^{-,f},
        q^f
        )$ for all $\theta,\bar\theta\in\R$ and deterministic flows $f$.
    Here, $F^{+,f}_{I\times E}\coloneqq (F^{+,f}_{i,e})_{i\in I,e\in E}$, $F^{-,f}_{E}\coloneqq (F^{-,f}_e)_{e\in E}$, and $q^f\coloneqq (q_e^f)_{e\in E}$ denote the cumulative edge in- and outflow functions, and queue length functions with respect to $f$.
\end{definition}

\begin{lemma}\label{lemma-sufficiently-continuous}
    A predictor $\predq_{i,e}$ that depends continuously on the cumulative inflow, total cumulative outflow and queue lengths, is $p$-continuous for any $p$ with $1< p < \infty$.
\end{lemma}
\begin{proof}
    The fact that $\predq_{i,e}(\emptyArg,\emptyArg, f)$ is continuous for every deterministic flow $f$ follows directly from the continuity of $\gamma_{i,e}$.
    Thus, it remains to show
    that $f^+\mapsto \predq_{i,e}(\emptyArg,\emptyArg,f)$ is sequentially weak-strong continuous from $L^p([0,M]\times D,\R_{\geq0})^{I\times E}$ to $C([0,M]\times D)$ for all $M$ and compact intervals $D$.

    \citet[Lemma~16]{CominettiCL15} showed that the maps $f_{i,e}^+\mapsto F_{i,e}^{+,f}$ and $f_e^+\mapsto q^f_e$ are sequentially weak-strong continuous from $L^p([0,M],\R_{\geq0})$ to $C([0,M],\R_{\geq 0})$ for any $M > 0$.
    Thus, the same holds for $f_e^+\mapsto (\theta\mapsto F_e^{+,f}(\theta - \transit_e) - q_e^f(\theta-\transit_e)) = F_e^{-,f}$.

    Let $(f^{+,k})_{k\in\N}$ be a sequence converging weakly to $f^+$ in $L^p([0, M])^{I\times E}$.
    Then, the sequence $(f^{+,k}_e)_{k\in\N}$ converges weakly to $f^+_e$ and $(f^{+,k}_{i,e})_{k\in\N}$ converges weakly to $f^+_{i,e}$ for every $e\in E$ and $i\in I$.
    By \cite{CominettiCL15}, the sequences $(F^{+,k}_{i,e})_{k\in\N}$, $(F^{-,k}_e)_{k\in\N}$ and $(q^k_e)_{k\in\N}$ converge strongly to $F_{i,e}^+$, $F_e^-$ and $q_e$, respectively, in $C([0,M])$ for all $e\in E$ and $i\in I$.
    This implies that these sequences also converge in $C(\R, \R_{\geq0})$ with respect to the topology induced by the uniform norm.

    We write $g^k \coloneqq (F^{+,k}_{I\times E}, F^{-,k}_E, q^k_E)$ for $k\in\N$ and $g\coloneqq (F^{+}_{I\times E}, F^{-}_E, q_E)$.
    The sequence $g^k$ converges to $g$ in $\mathcal G \coloneqq C(\R, \R_{\geq0})^{(I\times E)+E+ E}$ with respect to the uniform norm.
        We want to show that \begin{align*}
            \norm{
                \predq_{i,e}(\emptyArg,\emptyArg, f^k) - \predq_{i,e}(\emptyArg, \emptyArg, f)}_{C([0, M]\times D)
            }
            = \norm{
                \gamma_{i,e}(\emptyArg,\emptyArg, g^k) - \gamma_{i,e}( \emptyArg,\emptyArg, g)
            }_{C([0, M]\times D)}
        \end{align*}
        vanishes as $k$ approaches $\infty$.

    Let $\varepsilon>0$ be arbitrary.
    As $\gamma_{i,e}$ is continuous, for all $(\theta,\bar\theta)\in[0, M]\times D$ there exists some $\delta_{(\theta,\bar\theta)} > 0$ such that $\smallnorm{(\theta,\bar\theta, g) - (\theta',\bar\theta', g')}_\infty < \delta_{(\theta,\bar\theta)}$ implies\displayStyleForArxiv{\ensuremath{
            \smallabs{\gamma_{i,e}(\theta,\bar\theta, g) - \gamma_{i,e}(\theta', \bar\theta', g')} < \nicefrac{\varepsilon}{2}
        }}
    for $g'\in\mathcal G$.
    The compact set $[0,M]\times D$ is covered by\displayStyleForArxiv{\ensuremath{
            C\coloneqq \set{ B_{\delta_{(\theta,\bar\theta)}}(\theta,\bar\theta) | \theta\in[0,M],\bar\theta\in D}
        }} where $B_r(x)\coloneqq \set{y\in [0,M]\times D | \norm{y - x}_\infty < r}$ denotes the open ball in $[0,M]\times D$ around $x$ with radius $r$.
    By the compactness of $[0,M]\times D$ there is a subcover using finitely many of these open balls.
    Let $\delta$ be the minimum radius of these finitely many balls.
    There is $N\in\N$ with $\smallnorm{ g^k -  g }_\infty < \delta$ for all $k\geq N$.
    For arbitrary $\theta\in[0, M]$, $\bar\theta \in D$ and $k\geq N$ we infer\displayStyleForArxiv{\ensuremath{
            \smallabs{
                \gamma_{i,e}(\theta,\bar\theta, g^k) - \gamma_{i,e}( \theta,\bar\theta, g)
            }  < \nicefrac{\varepsilon}{2},
        }}
    which implies the desired result
    \(
    \smallnorm{
        \gamma_{i,e}(\emptyArg,\emptyArg, g^k) - \gamma_{i,e}( \emptyArg,\emptyArg, g)
    }_{C([0, M]\times D)} < \varepsilon.
    \)
\end{proof}

\subsection{Application Predictors}\label{predictors}\label{sec:predictors}
We now discuss several predictors and analyze whether the theorem above can be applied.
We begin with simple predictors and make them more sophisticated step-by-step.

The \emph{Zero-Predictor} predicts no queues for all times, i.e.,
\[
    \hat q^{\text{Z}}_{i,e}(\theta,\bar\theta,f) = 0.
\]
This predictor is trivially continuous, causal, and respects FIFO. In the resulting dynamic prediction flow, particles just always follow physically shortest paths.

The \emph{constant predictor} predicts that all queues will stay constant:
\[
    \hat q_{i,e}^{\text{C}}(\theta,\bar\theta, f)= q_e^f(\bar \theta).
\]
This leads to the mentioned special case of IDE flows. Since the constant predictor clearly is causal, depends continuously on the queue lengths and respects FIFO, we can apply \Cref{thm:existence} and, thus, reprove the existence of IDE flows shown in \cite{GrafHS20}.

The \emph{linear predictor} takes the derivative of the queues and extends them linearly up to some fixed time horizon $H \in \R_{\geq 0}\cup\{\infty\}$.
Formally it is defined as
\[
    \hat q_{i,e}^{\text{L}}(\theta,\bar\theta, f) \coloneqq
    \left( q^f_e(\bar \theta)+\partial_-q^f_e(\bar \theta)\cdot \min\{ \theta-\bar\theta, H \} \right)^+,
\]
where $(x)^+\coloneqq \max\{ x, 0 \}$ denotes the positive part of $x\in\R$.
The linear predictor is not in general continuous
since the partial derivative $\partial_-q_e(\bar \theta)$ might be discontinuous.

The \emph{regularized linear predictor} solves this by taking a rolling average of the past gradient with rolling horizon $\delta>0$ and extending the prediction according to this:
\[
    \hat q_{i,e}^{\text{RL}}(\theta,\bar\theta, f) \coloneqq
    \Big( q^f_e(\bar\theta) + \frac{q^f_e(\bar\theta) - q^f_e(\bar\theta - \delta)}{\delta} \cdot \min\{ \theta - \bar\theta, H \} \Big)^+
\]

    \begin{proposition}\label{prop:regPredictor}
        The regularized linear predictor is causal, respects FIFO and $p$-continuous for any $p>1$. It, thus, induces the existence of a DPE.
    \end{proposition}
    \begin{proof}
        It is clear that $\predq^\predRL$ is causal.
        Property~\eqref{eq:Cont-FlowDefProperties-OpAtCap} implies $q^f_e(\bar\theta) - q^f_e(\bar\theta - \delta) \leq \delta \cdot\capa_e$.
        Thus, $\predT^\predRL_{i,e}(\emptyArg,\bar\theta,f)$ is non-decreasing and $\predq^\predRL$ respects FIFO.
        Moreover, for any $g\in C(\R)$, the map $(x,f)\mapsto f(g(x))$ is continuous from $\R\times C(\R)$ to $\R$ such that we can apply~\Cref{lemma-sufficiently-continuous}.
    \end{proof}

\newcommand{\firstNN}[1]{\left\llbracket #1 \right\rrbracket}
The regularized linear predictor takes two samples of the past queue lengths (at times $\bar\theta$ and $\bar\theta-\delta$) and uses this information to predict future queue lengths up to the prediction horizon by a \piecewise\ linear function.
We generalize this idea by taking a fixed, finite set of observations of the past flow and employ a continuous transformation on these values to predict the future queues:
    In the following, we introduce a variant of this technique to embed machine-learning methods as a means to implement a causal, $p$-continuous predictor that respects FIFO.
Let $k_p$ be the number of past samples, $k_f$ the number of future samples, and $\delta > 0$ be some step size.
The arguments of the function $\gamma_{i,e}$ from \Cref{def:sufficient-regularity-condition} are mapped to $\R^d$ using an observation map $\sigma$.
This map returns $k_p$ samples of the queue length $q_{e'}^f$ and the edge load $L_{e'}^f\coloneqq F^{+,f}_{e'} - F^{-,f}_{e'}$ at the past times $\bar\theta-(i-1)\delta$ for every $i\in\firstN{k_p}$ and $e'\in E$.
Then, a continuous (machine-learned) transformation $\phi_{i,e}$ is applied resulting in interpolation points $\predq^\predMLRaw_{i,e}(\bar\theta+j\delta,\bar\theta,f)\geq 0$ for $j\in\firstN{k_f}$.
These two maps are depicted in the following:
\[
    \begin{tikzcd}[column sep=small]
\begin{array}{c}
            \theta, \bar\theta, \\
            F_{I\times E}^{+,f},F_{E}^{-,f},q^f
        \end{array}
        \arrow[r,mapsto, "\sigma"] &
        \begin{array}{c}
            \bar\theta,                                                      \\
            \left(q^f_{e'}(\bar\theta-(j-1)\delta)\right)_{\substack{e'\in E \\j\in\firstN{k_p}}},\\
            \left(L^f_{e'}(\bar\theta-(j-1)\delta)\right)_{\substack{e'\in E \\j\in\firstN{k_p}}}
        \end{array}
        \arrow[r,mapsto, "\phi_{i,e}"]
        &
        \left(\predq^\predMLRaw_{i,e}(\bar\theta+j\delta,\bar\theta,f)\right)_{\substack{j\in\firstN{k_f}}}
    \end{tikzcd}
\]
Finally, these points are linearly interpolated (in the first argument) to determine the value $\predq^\predMLRaw_{i,e}(\theta,\bar\theta,f)$.
This results in a causal, and $p$-continuous predictor.
However, $\predq^\predMLRaw$ could violate FIFO.
To prevent that, we apply the following post-processing:
We derive new interpolation points with $\predq^\predML_{i,e}(\bar\theta,\bar\theta, f) = q^f_{e}(\bar\theta)$ and \[
    \predq^\predML_{i,e}\left(\bar\theta + j\delta,\bar\theta, f \right)
    \coloneqq
    \max\left\{
    \begin{array}{c}
        \predq^\predMLRaw_{i,e}\left(\bar\theta + j\delta, \bar\theta, f \right), \\[.5em]
        \predq^\predML_{i,e}\left(\bar\theta + (j-1)\delta, \bar\theta, f \right) - \capa_e \delta
    \end{array}
    \right\}
\]
for all $j\in\firstN{k_f}$.
We extend the predictor constantly outside the interpolation points using
    $\predq^\predML_{i,e}\left(\theta,\bar\theta, f \right)=\predq^\predML_{i,e}\left(\bar\theta,\bar\theta, f \right)$ for
    $\theta<\bar\theta$ and
    $\predq^\predML_{i,e}\left(\theta,\bar\theta, f \right)=\predq^\predML_{i,e}\left(\bar\theta+k_f\delta,\bar\theta, f \right)$
    for $\theta>\bar\theta + k_f\delta$.
We call the resulting predictor $\predq^\predML_{i,e}$ the \emph{machine-learned predictor}.

\begin{proposition}
    If all maps $\phi_{i,e}$ are continuous and non-negative, the predictor $\hat q^{\text{ML}}$ is causal, respects FIFO, and depends continuously on the queue length functions.
    It, thus, induces the existence of a dynamic prediction equilibrium.
\end{proposition}
\begin{proof}
    As the observation map $\sigma$ only takes samples of the queues and the edge loads at times $\theta\leq \bar\theta$, the predictor $\predq^\predMLRaw$ and thus the predictor $\predq^\predML$ are clearly causal.

    To prove that $\predq^\predML$ respects FIFO, it suffices to show that $\predq^\predML_{i,e}(\theta_2, \bar\theta, f) - \predq^\predML_{i,e}(\theta_1, \bar\theta, f) \geq (\theta_2 - \theta_1)\cdot \capa_e$ holds for all $\theta_1\leq\theta_2$.
    This follows from the post-processing applied in the definition of $\predq^\predML_{i,e}$.

    With the same reasoning as for the regularized predictor, we observe, that the map $\sigma$ is continuous.
    Thus, the composition $\phi_{i,e}\comp\sigma$ is continuous as well.
    Hence, the mapping \[
        \begin{array}{c}
            \theta, \bar\theta, \\
            F_{I\times E}^{+,f},F_{E}^{-,f},q^f
        \end{array}
        \mapsto
        \begin{array}{c}
            \left( \predq^\predML_{i,e}(\bar\theta + j\delta, \bar\theta, f) \right)_{j=0,\dots,k_f}
        \end{array},
    \]
    that assigns the final interpolation points to the flow functions, is also continuous.
    Finally, evaluating the linear interpolation of these points at time $\theta$ is also a continuous operation.
\end{proof}

This leaves the question of how to choose the transformation maps in order to retrieve a good predictor.
In our experimental study, we use both a linear regression model and a simple neural network and train these models on previously computed flows.
We provide more details on the features and data used to train the predictor in~\Cref{subsec:machine-learned-predictors}.

Finally, the \emph{perfect predictor} predicts the queues exactly as they will evolve, i.e.
\[
    \hat q_{i,e}^{\text{P}}(\theta,\bar\theta, f) \coloneqq q^f_e(\theta).
\]
This predictor clearly is not causal and, thus, we can not apply our existence result here. However, dynamic predicted flows with respect to this predictor do exist as those are just dynamic Nash equilibria for which existence has been proven in \cite{CominettiCL15}.

 \section{Approximate Dynamic Prediction Equilibria}

We augment the introduction of DPE by an analysis of the computation of an approximate variant of DPE.
We recall that in a DPE, flow is only sent along predicted shortest paths.
More specifically, flow enters an edge only if the predicted delay $\predDel_{i,e}(\theta,\theta,f)$ vanishes.
This inspires the following definition of approximate DPE, in which agents enter an edge only if the predicted delay is small enough.

\begin{definition}
    A pair $(\predq, f)$ of predictors $\predq$ and a flow over time $f$ is called an \emph{$\varepsilon$-approximate DPE} up to time $H\in\R_{>0}\cup\{\infty\}$, if $f$ is feasible up to time $H$ and for all $i\in I$, $e\in E$, and $\theta< H$ it holds that \[
        f_{i,e}^+(\theta) > 0
        \implies
        e\in E_i \,\land\,
        \predDel_{i,e}(\theta, \theta, f) \leq \varepsilon.
    \]
\end{definition}

The goal of this section is to develop an algorithm for computing $\varepsilon$-approximate DPE provided that all predictors are causal, respect FIFO, fulfill a regularity condition, and are computable as piecewise linear functions.
We start off with an algorithm that computes so-called \emph{$\delta$-routed DPE} in which route updates happen only at discrete time steps of size $\delta > 0$.
After an outline of the algorithm in \Cref{sec:outline-algo}, \Cref{sec:compute-outflow-rates} points out how to compute outflow rates and queue lengths given piecewise constant inflow rates.
In \Cref{sec:correctness-termination} we prove the correctness and termination of the algorithm.
Finally, \Cref{sec:delta-dpe-are-approx-dpe} shows that by imposing a Lipschitz-condition on the predictors, any $\delta$-routed equilibrium flow is also an $\varepsilon$-approximate DPE where $\varepsilon$ depends linearly on $\delta$.

\subsection{Extension-Based Approximation Algorithm}\label{sec:outline-algo}

As agents in a DPE behave like infinitesimal particles in a continuous flow, the routes chosen may change indefinitely often.
For computational analyses it is infeasible to calculate shortest paths this often.
Hence, we discretize the points in time at which routes are updated.

More specifically, shortest paths are recalculated every $\delta$ time units for some $\delta > 0$.
This means that particles arriving at an intermediate node $v$ at some time $\theta$ will use a predicted shortest path computed at time\displayStyleForArxiv{\ensuremath{
        \vartheta_\delta(\theta) \coloneqq \floor{\frac{\theta}{\delta}}\cdot \delta,
    }}
leading to the following variant of DPE whose computation we will focus on for the moment.

\begin{definition}
    Let $f$ be a dynamic flow, $\predq=(\predq_{i,e})_{i,e}$ a set of predictors and $\delta > 0$.
    The pair $(f, \predq)$ is called a \emph{$\delta$-routed DPE} up to time $H\in\R \cup\{\infty\}$ if $f$ is feasible up to time $H$ and for all $e\in E$, $i\in I$ and almost all $\theta < H$ it holds that \[
        \infl_{i,e}(\theta) > 0 \,\implies\, e \in \predE_{i}\left(\vartheta_\delta(\theta),\vartheta_\delta(\theta), f\right).
    \]
\end{definition}

    In other words, whenever there is positive inflow into an edge, that edge must be active at time $k\cdot \delta$ as predicted at time $k\cdot \delta$ with $k\cdot\delta\leq\theta < (k+1)\cdot \delta$.

We require that network inflow rates are piecewise constant with finitely many jumps.
This enables us to find a $\delta$-routed DPE with \piecewise\ constant edge inflow and outflow rates.
As a result, cumulative edge inflow and outflow functions, queue length and exit time functions of edges and paths, as well as earliest arrival time functions are \piecewise\ linear.
Consequently, we assume that the predicted queue length functions $\predq_{i,e}(\emptyArg, \bar\theta, f)$ of all predictors are \piecewise\ linear.
    We formally specify these classes of functions as follows.

    \begin{definition}
        A function $f: \R \to \R$ is called \emph{(right) \piecewise\ constant} if there exists a finite chain $\xi_1 < \xi_2 < \cdots < \xi_k$ with $\xi_i\in \R$ such that $f$ is constant on $(-\infty, \xi_1)$, on $[\xi_i, \xi_{i+1})$ for every $i\in\firstN{k-1}$, and on $[\xi_k, \infty)$.
    \end{definition}

    \begin{definition}
        A continuous function $f: \R\to\R$ is called \emph{\piecewise\ linear} if there exists a finite chain $\xi_1 < \xi_2 < \cdots < \xi_k$ with $\xi_i\in\R$ such that $f$ is affine on $(-\infty, \xi_1)$, on $[\xi_i, \xi_{i+1}]$ for every $i\in\firstN{k-1}$, and on $[\xi_k, \infty)$.
    \end{definition}

The predictors in our simulation must be causal and respect FIFO.
The first requirement is necessary, as we build our flow in an extension based approach, so that predictions cannot rely upon the future evolution of the flow.
The second allows us to use shortest path algorithms for FIFO cost functions.

In the following we give an overview on the computation of a $\delta$-routed DPE.
Initially, we begin with a ``zero''-flow $f$ with $\infl,\outfl\colonequiv 0$ as our $\delta$-routed DPE flow up to time $H=0$. We call $H$ the \emph{flow horizon} of $f$.
We aim to compute a $\delta$-routed DPE flow up to some $\Hcomp\in\R$.
The algorithm consists of two different phases that need to be repeated multiple times:
A \emph{routing phase} and a \emph{distribution phase}.

A \emph{routing phase} is run when the $\delta$-routed DPE flow $f$ has been calculated up to some multiple $H = k\cdot \delta$ of $\delta$.
All routes that have been calculated up to time $H$ are invalidated, and new routes are determined.
This is done in the following two steps:

\begin{enumerate}[label=(R\arabic*)]
    \item Gather predictions $\predq_{i,e}(\emptyArg, H, f)$ as \piecewise\ linear functions for all commodities $i\in I$ and edges $e\in E$.
    \item Compute the set of active outgoing edges $\outEdges{v}\cap \predE_{i}(H, H, f)$ for all nodes $v\in V_i$ and commodities $i\in I$.
\end{enumerate}

In the \emph{distribution phase} the dynamic flow gets extended.
Assume we have computed a $\delta$-routed DPE flow $f$ up to flow horizon $H$ and want to extend $H$ up to some time $H' > H$.
To ensure flow conservation, we distribute the \emph{node inflow rate} \[
        b_{i,v}^-(\theta) \coloneqq \sum_{e\in\inEdges{v}} \outfl_{i,e}(\theta) + \CharF{v=s_i}\cdot u_i(\theta)
    \]
of a node $v\in V_i$ to outgoing edges.
A distribution phase then consists of the following  two steps.
\begin{enumerate}[label=(D\arabic*)]
    \item\label{step:distribute} Update $f$ with the deterministic flow with respect to the updated inflow rates \[
        \restrict{f_{i,e}^+}{[H, \infty)}
        \colonequiv
        \begin{cases}
            \frac{b^-_{i,v}(H)}{\abs{\outEdges{v} \cap \predE_i(\vartheta_\delta(H),\vartheta_\delta(H), f)}}, & \text{if $e\in \predE_i(\vartheta_\delta(H),\vartheta_\delta(H), f)$,} \\
            0,                                                                                                 & \text{otherwise,}
        \end{cases}
    \] for all commodities $i\in I$ and edges $e=vw\in E$.
    \item\label{step:max-H'} Determine the maximal $H'\leq \vartheta_\delta(H)+\delta$ such that $b_{i,v}^-$ is constant on $[H, H')$ (according to the updated deterministic flow) for every node $v$ and commodity $i\in I$,
    and set $H\coloneqq H'$.
\end{enumerate}
Here, $\vartheta_\delta(H) + \delta$ denotes the beginning of the next routing phase based on $H$.

The distribution phase is executed multiple times until the flow horizon $H$ reaches the next multiple of $\delta$; then a new routing phase is initiated.
Once the flow horizon $H$ reaches the computation target $\Hcomp$ we terminate the process and return the computed flow $f$.
A schematic overview of this algorithm is depicted in \Cref{fig:schem-overview-algo-dpes}.

\begin{figure}
    \centering
    \begin{tikzpicture}[font=\small]
    \node[draw,
        rounded corners=0.2cm,
        minimum width=3.5cm,
        text width=3.5cm,
        text centered,
        minimum height=1cm] (init) {
            Initial flow $f \equiv 0$\\with horizon $H = 0$
        };

    \node[draw,
        rounded corners=0.2cm,
    below=0.3cm of init,
    minimum width=3.5cm,
    minimum height=1cm,
    text centered
    ] (route) {\makebox{{\begin{varwidth}{\linewidth}
        {\centering\textbf{Routing Phase}\\}
        (R1) Gather $(\predq_{i,e}(\emptyArg, H,f))_{i,e}$\\
        (R2) Compute $(\predE_i(H, H, f))_{i}$
    \end{varwidth}}}};

    \node[draw,
        rounded corners=0.2cm,
        below=0.3cm of route,
        minimum width=3.5cm,
        minimum height=1cm,
        text centered
    ] (distribute) {\makebox{{\begin{varwidth}{\linewidth}
        {\centering\textbf{Distribution Phase}\\}
        (D1) Distribute $b_{i,v}^-(H)$ uniformly to \\
        \phantom{(D2) }$\outEdges{v}\cap \predE_{i}(\vartheta_\delta(H),\vartheta_\delta(H), f)$
        on $[H, \infty)$\\
        (D2) Determine maximal $H'\leq \vartheta_\delta(H)+\delta$ such that\\
        \phantom{(D1) }$(b_{i,v}^-)_{i,v}$ is constant on $[H, H')$, and set $H\coloneqq H'$
    \end{varwidth}}}};
     
    \node[draw,
        diamond,
        above right=-0.63cm and 2cm of distribute,
        minimum width=2.5cm,
        aspect=1.5,
        inner sep=0] (checkTerminate) {
            $H\geq \Hcomp$
    };
    \node[draw,
        diamond,
        above=.7cm of checkTerminate,
        minimum width=2.5cm,
        aspect=1.5,
        inner sep=0] (checkReroute) {
            $H = \vartheta_\varepsilon(H)$
        };
    
    \node[draw,
        rounded corners=0.2cm,
        minimum width=2.5cm,
        text centered,
        minimum height=0.5cm,
        below=0.7cm of checkTerminate] (return) {
        Return $f$
    };
\coordinate[xshift=-0.05857864376cm,yshift=-0.05857864376cm] (distributeNorthEast) at (distribute.north east);
    \draw[-latex] (init) edge (route) (route) edge (distribute);
    \draw[-latex] (distribute) edge (checkTerminate.west);
    \draw[-latex] (checkTerminate) edge node[pos=0.4,fill=white,inner sep=2pt]{No}(checkReroute);
    \draw[-latex] (checkReroute.west) -- node[pos=0.4,fill=white,inner sep=2pt]{Yes}(route.east);
    \draw[-latex] (checkReroute) edge node[pos=0.4,fill=white,inner sep=2pt]{No}(distributeNorthEast);
    \draw[-latex] (checkTerminate) edge node[pos=0.4,fill=white,inner sep=2pt]{Yes}(return);
\end{tikzpicture}     \caption{A schematic overview of the computation of a $\delta$-routed DPE.}
    \label{fig:schem-overview-algo-dpes}
\end{figure}

\subsubsection*{Remarks on the Implementation}

There are three major aspects to consider when implementing the algorithm.
Firstly, while it is cheap to compute the predicted queue length functions $\predq_{i,e}(\emptyArg, H, f)$ for the different causal predictors introduced in \Cref{predictors}, it is computationally expensive to compute the set of active edges $\predE_{i}(H,H,f)$.
Secondly, a difficulty in the distribution phase lies in computing the outflow rates when extending the inflow rates with a new constant as in Step~\ref{step:distribute}.
Thirdly, it is unclear how to find the maximal time $H'$ such that $b_{i,v}^-$ is constant on $[H, H')$ quickly.
It is a good idea to solve the latter two problems in combination, as any change to an edge outflow rate $\outfl_{i,vw}$ (induced by changing edge inflow rates) results in a change of the target node's inflow rate $b_{i,w}^-$.
The solution to these two problems is discussed in \Cref{sec:compute-outflow-rates,sec:correctness-termination}.
Additionally, we reduce the work in consecutive distribution phases by only considering nodes $v$ whose inflow rates $(b_{i,v}^-)_i$ changed at time $H$.

Computing the active edges $\predE_i(H, H, f)$ can be solved by a dynamic variant of the Bellman-Ford algorithm covered by \cite{Bolin08}.
However, because of its poor running time, it is beneficial to use a Dijkstra-based algorithm.
With this approach we have to calculate the active outgoing edges for each node and commodity with a separate run of the algorithm.
Therefore, we defer this calculation until we notice a node $v$ in the distribution phase with positive inflow rate $b_{i,v}^-$ and only calculate the active outgoing edges of $v$ on demand.
In most cases, only a small subset of nodes will experience inflow of any given commodity; in our experiments this proved to be much faster than the Dynamic Bellman-Ford algorithm.

Nevertheless, for predictors $\predq_i$ that are constant in the first argument, i.e. if $\predq_{i,e}(\emptyArg, \bar\theta, f)$ is constant for all $e\in E$, $\bar\theta\in \R$ and deterministic flows $f$, a single run of a simple static version of Dijkstra's algorithm suffices to compute the set of all active edges.
For predictors of this type, e.g., the Zero-predictor and the constant predictor, this approach is the preferred method to compute $\predE_i(H,H,f)$.

An implementation of this algorithm using Python is publicly available in \cite{GitHub2021}.

\subsubsection*{Adjustments for Computing IDE flows}

In Step~\ref{step:distribute} we choose to distribute the node inflow rate  equally to all active outgoing edges.
Of course, this does not ensure, that edges that were active at time $\vartheta(H)$ will remain active throughout the interval $(H, H')$.
In the case of instantaneous dynamic equilibria, that is, if all commodities use the constant predictor, and if we have only a single sink, we can use the more sophisticated so-called \emph{water filling algorithm} (c.f. \cite[Chapter~5]{Graf2024}) to distribute the node inflow to the active outgoing edges.
This algorithm ensures that edges remain active for some time period.

When computing a $\delta$-routed DPE we usually deal with arbitrary predictors, and hence use the heuristic to distribute incoming flow uniformly across the active outgoing edges.
However, the adjustments necessary to compute exact IDE flows are modest:
In the routing phase, we can use a static version of Dijkstra's algorithm to compute the instantaneous active edges $(E_{i, H})$ for some time $H$.
We then compute the maximal $H'$ such that no node inflow rate changes and the set of active edges remains the same on $(H, H')$.
Then, we distribute the node inflow rate according to the water filling algorithm.
We repeat this procedure until we have reached the targeted flow horizon $\Hcomp$.

For further analysis of the resulting ``natural extension algorithm'' the reader is referred to \cite{Graf2024}.
A key result therein is that there is an implementation of this algorithm which is guaranteed to terminate.
Similarly, in \Cref{thm:termination} we prove that our proposed extension procedure terminates under mild conditions on the predictors.

\subsection{Outflow Rates of Piecewise Constant Inflow Rates}\label{sec:compute-outflow-rates}
To solve the problems described above, we first consider a few theoretical observations involving the computation of general deterministic dynamic flows.
The idea of the introduced algorithm is to utilize the simple structure of flows over time generated by \piecewise\ constant inflow rates.
This is demonstrated by the following theorem.

{
\renewcommand{\phi}{H}
\begin{theorem}\label{thm:extend-edge}
    Let $f$ be a deterministic flow, let $(g_{i,e})_{i\in I}\in\R_{\geq0}^{I}$ be a set of new constant inflow rates into an edge $e$ beginning from time $\phi\in\R$, and let $g_e \coloneqq \sum_{i\in I} g_{i,e}$.
    Let $h=(\outfl[h],\infl[h])$ denote the deterministic flow with inflow rates \[
        h_{i,e}^+(\theta) \coloneqq \begin{cases}
            f_{i,e}^+(\theta), & \text{for $\theta < \phi$,}    \\
            g_{i,e},           & \text{for $\theta \geq \phi$,}
        \end{cases}
    \]
    Then $\outfl[h]_{i,e}$ is given by $\restrict{\outfl[h]_{i,e}}{(-\infty,T^f_e(\phi))} \aequal \restrict{\outfl_{i,e}}{(-\infty, T^f_e(\phi))}$ and by the following cases:
    \begin{enumerate}[label=(C\arabic*):, ref=(C\arabic*)]
        \item\label{case-1} $g_e = 0$. Then $\restrict{\outfl[h]_{i,e}}{[T^f_e(\phi), \infty)} \aequiv 0$.

        \item\label{case-2} $g_e > 0 \,\land\, \left( q^f_e(\phi) = 0 \,\lor\, g_e \geq \capa_e \right)$. Then $\restrict{\outfl[h]_{i,e}}{[T^f_e(\phi), \infty)} \aequiv \min\{\capa_e, g_e\}\cdot \frac{g_{i,e}}{g_e}$.

        \item\label{case-3} $g_e \in (0, \capa_e) \,\land\, q^f_e(\phi) > 0$. Then, with $T_{\depl} \coloneqq \phi + \frac{q^f_e(\phi)}{\capa_e - g_e}$,
        \begin{align*}
            \restrict{\outfl[h]_{i,e}}{[T^f_e(\phi), T_{\depl} + \transit_e)}
            \aequiv \capa_e \cdot \frac{g_{i,e}}{g_e}
            \quad\text{and}\quad
            \restrict{\outfl[h]_{i,e}}{[T_{\depl} + \transit_e, \infty)}
            \aequiv g_{i,e}.
        \end{align*}
    \end{enumerate}
\end{theorem}
\begin{proof}
    As the inflow rates of $f$ and $h$ into edge $e$ match up to time $\phi$ (and $f$ and $h$ are deterministic), the queue and exit time functions of edge $e$ coincide up to time $\phi$, and the outflow rate functions match up to time $T_e(\phi)$.
    For $\theta > T_e(\phi)$, we handle all three cases individually and utilize the observation\displayStyleForArxiv{\ensuremath{
            q_e^h(\theta - \transit_e) = q_e(\phi) + g_e \cdot (\theta - \transit_e - \phi) - \int_{\phi + \transit_e}^{\theta} \outfl[h]_e\diff\leb.
        }}

    \ref{case-1}. From $g_e = 0$ and \eqref{eq:Cont-FlowDefProperties-OpAtCap} it follows \(
    q_e^h(\theta-\transit_e)
    \leq q_e(\phi) - \int_{\phi + \transit_e}^{T_e(\phi)} \outfl[h]_e\diff\leb = 0
    \)
    for all $\theta > T_e(\phi)$.
    Applying~\eqref{eq:Cont-FlowDefProperties-OpAtCap} again yields $\outfl[h]_{i,e}(\theta) = 0$ for $\theta > T_e(\phi)$.

    \ref{case-2}. We show $\outfl[h]_e(\theta) = \min\{\capa_e, g_e\}$ for almost all $\theta > T_e(\phi)$.
    Then, since any $\xi$ fulfilling $T_e^h(\xi) = \theta$ is greater than $\phi$, property~\eqref{eq:FIFO} leads to the conclusion $\outfl[h]_{i,e}(\theta) = \min\{\capa_e, g_e\}\cdot \frac{g_{i,e}}{g_e}$ for almost all $\theta > T_e(\theta)$.

    In the case $g_e > \capa_e$, for any $\theta > T_e(\phi)$, the queue length fulfills $q_e^h(\theta - \transit_e)> 0$.
    Therefore $\outfl[h]_{e}(\theta) = \capa_e$ holds for almost all $\theta > T_e(\phi)$.
    For $g_e = \capa_e$ and $q_e(H) > 0$, the same reasoning applies.
    Now assume $g_e\in(0,\capa_e)$ and $q_e(\phi) = 0$.
    For any $\theta > T_e(\phi)$, we have\displayStyleForArxiv{\ensuremath{
            q^h_e(\theta - \transit_e) = \max_{\xi\in[\phi, \theta - \transit_e]} \int_\xi^{\theta-\transit_e} \infl[h]_e - \capa_e \diff\leb = 0.
        }}
    Constraint \eqref{eq:Cont-FlowDefProperties-OpAtCap} yields $\outfl[h]_{e}(\theta) = g_e$ for almost all $\theta > T_e(\phi)$.

    \ref{case-3}. It suffices to prove $\restrict{\outfl[h]_e}{(T_e(\phi), T_{\depl} + \transit_e)} \aequiv \capa_e$ and $\restrict{\outfl[h]_e}{(T_{\depl} + \transit_e, \infty)} \aequiv g_e$.
    For $\theta \in (T_e(\phi), T_{\depl} + \transit_e)$ we have\displayStyleForArxiv{\ensuremath{
        q_e^h(\theta - \transit_e) \geq q_e(\phi) - (\capa_e - g_e)\cdot (\theta - \transit_e - \phi) > 0,}}
    implying $\outfl[h]_e(\theta) = \capa_e$ for almost all of these $\theta$.
    Therefore, $T_{\depl}$ fulfills $q_e^h(T_{\depl}) = q_e^h(\phi) - (\capa_e - g_e)\cdot (T_{\depl} - \phi) = 0$.
    For $\theta > T_{\depl} + \transit_e$, we compute \displayStyleForArxiv{\ensuremath{
        q_e^h(\theta-\transit_e) = \max_{\xi\in [T_{\depl}, \theta-\transit_e]} \int_\xi^{\theta-\transit_e} \infl[h]_e - \capa_e \diff\leb = 0,
    }}
    implying $\outfl[h]_e(\theta) = g_e$. \qedhere
\end{proof}

\begin{remark}\label{rem:extend-queues}
    The proof also describes the evolution of the queue length starting at time $\phi$.
    This means, in the setting of \Cref{thm:extend-edge}, the queue $q_e^h$ is given for $\theta \geq \phi$ as follows.
    In the cases~\ref{case-1} and~\ref{case-3} we have
    \[
        q^h_e(\theta) = \begin{cases}
            q^f_e(\phi) - (\theta - \phi)\cdot (\capa_e - g_e), & \text{for $\theta \in [\phi, T_{\depl}]$,} \\
            0,                                                  & \text{for $\theta \geq T_{\depl}$,}
        \end{cases}
    \]
    where $T_{\depl} \coloneqq \phi + \frac{q^f_e(\phi)}{\capa_e - g_e}$ denotes the depletion time.
    For~\ref{case-2}, no queue depletion occurs and for $\theta \geq \phi$ we have\displayStyleForArxiv{\ensuremath{
            q_e(\theta) = q_e(\phi) + (\theta - \phi)\cdot \max\{g_e - \capa_e, 0\}.
        }}
\end{remark}

\subsection{Correctness and Termination}\label{sec:correctness-termination}

The previous section allows us to calculate the deterministic outflow rates for the case that new constant inflow rates are assigned to an edge as in Step~\ref{step:distribute}:
In the implementation, we maintain \piecewise\ constant functions $\infl_{i,e}$ and $\outfl_{i,e}$ for all $i\in I$ and $e\in E$, as well as \piecewise\ linear functions $q_e$ for all $e\in E$.
Before and after each distribution phase, these functions represent a deterministic flow that is feasible up to time $H$.
In Step~\ref{step:distribute} the inflow rates are then updated starting from time $H$ and the outflow rates as well as the queue length functions are updated according to \Cref{thm:extend-edge} and \Cref{rem:extend-queues} starting from times $T_e(H)$ and $H$, respectively, while preserving the deterministic property of $f$.

To detect the maximum $H'$ such that $b_{i,v}^-$ stays constant on $[H, H')$ for all $i\in I$, $v\in V$ in Step~\ref{step:max-H'}, we employ a priority queue that contains all events of the form
\begin{enumerate}[label=(E\arabic*)]
    \item\label{ev:outflow-rate} ``The outflow rate $\outfl_{i,e}$ changes at time $\theta$'' for all $i\in I$, $e\in E$ and $\theta > H$, and
    \item\label{ev:net-inflow} ``The network inflow rate $u_i$ changes at time $\theta$'' for all $i\in I$ and $\theta > H$.
\end{enumerate}

To achieve this, whenever we update an edge's outflow rate $\outfl_{i,e}$ in Step~\ref{step:distribute}, we generate corresponding events of type \ref{ev:outflow-rate} and possibly remove existing events, that are rendered invalid by the update.
As we require $\transit_e > 0$, the added events happen after $H$.
The events of type~\ref{ev:net-inflow} are enqueued in the initialization of the algorithm.

A change in a node inflow rate $b_{i,v}^-$ requires that either the network inflow rate has changed (and $v=s_i$) or that an outflow rate of an incoming edge has changed.
Thus, the minimum time $\theta$ of all events in the priority queue is a lower bound to the maximum time $H'$ such that $\restrict{b_{i,v}^-}{[H, H')}$ is constant for all $i\in I$ and $v\in V$.
If there is only a single event in the queue at time $\theta$, then $b_{i,v}^-$ changes at time $\theta$, and we have $H' = \theta$.
If multiple events occur at time $\theta$ and their changes to $b_{i,v}^-$ balance out for all $i\in I$ and $v\in V$, we can simply ignore these events, because this implies that all $b_{i,v}^-$ stay constant with value $b_{i,v}^-(H)$ for an even longer period than up to time $\theta$.
Otherwise, $b_{i,v}^-$ changes at time $\theta = H'$.
Once $H'$ is determined, we remove all events with time $\theta \leq H'$ from the priority queue.

We now show the correctness and termination of the proposed algorithm.

\begin{definition}
    A predictor $\predq_{i,e}$ is \emph{computable as \piecewise\ linear functions}, if the function $\predq_{i,e}(\emptyArg,\bar\theta, f)$ is \piecewise\ linear and there exists an algorithm computing this function for every $\bar\theta\in\R$ and deterministic flow $f$ with \piecewise\ constant inflow rates.
\end{definition}

\begin{theorem}\label{thm:termination}
    If all predictors are causal, computable as \piecewise\ linear functions and respect FIFO, then the algorithm described in \Cref{sec:outline-algo} returns a $\delta$-routed DPE up to time $\Hcomp\in\R$ in finite time.
\end{theorem}
\begin{proof}
    If the algorithm terminates, the returned flow is feasible up to time $\Hcomp$.
    The fact that at any time $\theta$ we extend the flow along edges in $\predE_{i}(\vartheta_\delta(\theta),\vartheta_\delta(\theta),f)$ implies that $(\predq, f)$ is a $\delta$-routed DPE up to time $\Hcomp$.

    To ensure that the algorithm terminates, we note that a single change in the node inflow rate $b_{i,v}^-$ at some time $\theta$ can cause finitely many events of type~\ref{ev:outflow-rate} that all happen no earlier than at time $\theta + \mintransit > \theta$.
    Assume, we have computed a flow up to time $H$.
    Then, all events that occur up to time $H+\nicefrac{\mintransit}{2}$ have already been enqueued.
    After processing these finitely many events, we will have computed a flow up to time $H+\nicefrac{\mintransit}{2}$.
    This suffices to assess that the algorithm terminates.
\end{proof}

\begin{remark}
    For a more thorough analysis, let $\maxoutdeg \coloneqq \max_{v\in V}\abs{\outEdges{v}}$ denote the maximum out-degree.
    Let $P_i$ denote the number of jumps of $u_i$ and $P\coloneqq \sum_{i\in I}P_i$.
    Then, during the interval $[0, \mintransit)$, at most $k_0 \coloneqq P$ events can happen.
    A single event that occurs at time $\theta$ can cause at most $2 \maxoutdeg \cdot \abs{I}$ events all of which happen later than $\theta + \mintransit$.
    Thus, the number of events that occur within the interval $[\mintransit, 2\mintransit)$ is bounded from above by $k_1\coloneqq k_0 \cdot (1 + 2\maxoutdeg\cdot \abs{I})$.
    Similarly, for any $l \in \N$ the number of events that happen in the interval $[l\mintransit, (l+1)\mintransit)$ is bounded by $k_l\coloneqq k_0\cdot (1+2\maxoutdeg\cdot \abs{I})^l$.
    Utilizing the geometric sum, the number of events that occur up to time $\ceil{\Hcomp / \mintransit} \cdot \mintransit \geq \Hcomp$ is upper bounded by \[
        \sum_{l=0}^{\ceil{\frac{\Hcomp}{\mintransit}}} k_l = P \cdot \frac{ \left(1 + 2\maxoutdeg\cdot \abs{I} \right)^{\ceil{\frac{\Hcomp}{\mintransit}}+1} - 1}{2\maxoutdeg \cdot \abs{I}},
    \]
    and thus the number of distribution phases is ${\bigO}{( P\cdot (1 + 2\maxoutdeg\cdot \abs{I})^{\nicefrac{\Hcomp}{\mintransit}+2})}$.
\end{remark}

\subsection{Computing Approximate Dynamic Prediction Equilibria}\label{sec:delta-dpe-are-approx-dpe}

Finally, we show that if the predictors fulfill a Lipschitz condition, any $\delta$-routed DPE is also an $\varepsilon$-approximate DPE.
The exact notion of the Lipschitz condition is the following.

\begin{definition}
    A predictor $\predq_{i,e}$ is called \emph{$L$-Lipschitz} with $L\in[0, \infty)$ if for every feasible flow $f$ the function $\predq_{i,e}(\emptyArg,\emptyArg,f)$
        fulfills the Lipschitz-condition with respect to the constant $L$, i.e. \[
        \forall\, (\theta,\bar\theta),(\theta',\bar\theta')\in\R^2:
        \quad
        \abs{\predq_{i,e}(\theta,\bar\theta, f) - \predq_{i,e}(\theta',\bar\theta', f)}\leq L\cdot\norm{(\theta,\bar\theta)-(\theta',\bar\theta')}_\infty.
    \]
\end{definition}

\begin{theorem}\label{thm:approximation-quality}
    If $(\predq, f)$ is a $\delta$-routed DPE and all predictors are $L$-Lipschitz, then $(\predq, f)$ is an $\varepsilon$-approximated DPE with $\varepsilon\coloneqq (\nicefrac{L}{\capa_{\min}}+2)\cdot(\nicefrac{L}{\capa_{\min}}+1)^{\abs{V}-1}\cdot\delta$ and $\capa_{\min}\coloneqq \min_{e\in E} \capa_e$.
\end{theorem}
\begin{proof}
    Using the triangle inequality and the Lipschitz-property of $\predq_{i,e}$, we deduce that the function $\predT_{i,e}(\emptyArg, \emptyArg, f)$ is $(\nicefrac{L}{\capa_e} + 1)$-Lipschitz.
    Hence, for any simple path $P=e_1\cdots e_k$, the function $\predT_{i,P}(\emptyArg, \emptyArg, f)$ is Lipschitz-continuous with constant $(\nicefrac{L}{\capa_{\min}} +1)^k\leq (\nicefrac{L}{\capa_{\min}} +1)^{\abs{V}-1}$.
    The pointwise minimum $\predl_{i,v}(\emptyArg,\emptyArg,f)$ of finitely many such functions is  $(\nicefrac{L}{\capa_{\min}} +1)^{\abs{V}-1}$-Lipschitz.
    By applying the triangle inequality once more, we see that $\predDel_{i,e}(\emptyArg, \emptyArg, f)$ is Lipschitz-continuous with constant $(\nicefrac{L}{\capa_{\min}} +2)\cdot (\nicefrac{L}{\capa_{\min}}+1)^{\abs{V}-1}$.
    
    Now, let $\theta\in\R$ be arbitrary and assume\displayStyleForArxiv{\ensuremath{
            \predDel_{i,e}\left(\floor{\theta/\delta}\cdot\delta,\floor{\theta/\delta}\cdot\delta,f\right) = 0.
        }}
    This implies \begin{align*}
        \predDel_{i,e}(\theta,\theta, f)
         & = \abs{\predDel_{i,e}(\theta, \theta, f) - \predDel_{i,e}\left(\floor{\theta/\delta}\cdot\delta,\floor{\theta/\delta}\cdot\delta,f\right) } \\
         & \leq (\nicefrac{L}{\capa_{\min}}+2)\cdot(\nicefrac{L}{\capa_{\min}}+1)^{\abs{V}-1}\cdot\delta = \varepsilon.
    \end{align*}
    Therefore, as $(\predq, f)$ is a $\delta$-routed DPE, it is also an $\varepsilon$-approximated DPE.
\end{proof}

\begin{remark}
    Note, that the approximation guarantee of \Cref{thm:approximation-quality} degrades exponentially with the number of nodes in a network.
    There exist networks in which also the achieved approximation quality $\varepsilon$ degrades exponentially:
    Assume that the network consists of a chain $P$ of edges from node $s$ to node $t$ and the predictor for each edge $e$ of the chain is given by $\hat q_{i,e}(\theta,\bar\theta,f)\coloneqq \theta\cdot L$, and let $\tau_e=0$ and $\nu_e=1$.
    Then, the predicted exit time of this chain $P$ is $\hat T_{i,P}(\theta,\bar\theta,f)=\theta\cdot (L+1)^{\abs V -1}$.
    We add an additional edge $e^*$ from $s$ to $t$ with $\hat q_{i,e^*}(\theta,\bar\theta, f)\coloneqq 0$ and $\tau_{e^*} \coloneqq 0$.
    Now assume a network inflow rate of $u \coloneqq \CharF{[0,1)}$.
    A flow $f$ which sends all particles along the chain is a $\delta$-routed DPE with $\delta=1$.
    However, the minimal $\varepsilon$ such that $f$ is an $\varepsilon$-approximated DPE is $\varepsilon = (L+1)^{\abs V-1}$.

    Note, that this does not imply that other methods of finding $\varepsilon$-approximated DPE must also suffer from this lower bound.
\end{remark}

    We shortly discuss whether the introduced predictors are $L$-Lipschitz.
    Clearly, the Zero-Predictor $\predq_{i,e}^\predZ$ is $0$-Lipschitz.
    The linear predictor, on the other hand, is not Lipschitz: Whenever $\partial_- q_e(\bar\theta)$ jumps, we cannot reduce the prediction difference by moving $\bar\theta'$ closer to $\bar\theta$.
    The same holds for the regularized linear predictor with infinite prediction horizon $H$: Assume that the difference quotient at time $\bar\theta$ is positive and slightly different to the one at time $\bar\theta'$.
    Then, the prediction difference is not bounded, as we can always predict at times $\theta=\theta'$ further in the future resulting in a higher prediction difference.
    On the positive side, we can show the Lipschitz-condition for several predictors--including the regularized linear predictor with \emph{finite} prediction horizon--by utilizing the following proposition:
    \begin{proposition}\label{queue-lipschitz}
        If $f$ is a feasible flow with bounded network inflow rates, i.e. $\norm{u_i}_\infty < \infty$, the queue length function $q_e^f$ of an edge $e=vw$ is $L_{q_e}$-Lipschitz with
        \[
            L_{q_e} \coloneqq \max\{ \capa_e, \sum_{i\in I}\CharF{v=s_i}\norm{u_i}_\infty + \sum_{e'\in\inEdges{v}} \capa_{e'}\}.
        \]
    \end{proposition}
    \begin{proof}
        This directly follows from the flow conservation constraint together with the queue dynamic $q_e^f(\theta) = \int_0^\theta f_e^+(z) - f_e^-(z + \transit_e) \diff z$.
    \end{proof}

    \begin{corollary}
        If all network inflow rates are bounded, then the constant predictor $\predq^\predC_{i,e}$ and the perfect predictor $\predq^\predP_{i,e}$ are $L_{q_e}$-Lipschitz.
        Moreover, if the prediction horizon $H$ of the regularized linear predictor $\predq^\predRL_{i,e}$ is finite, then $\predq^\predRL_{i,e}$ is $(2 L_{q_e} (2+\nicefrac H\delta))$-Lipschitz where $\delta$ is the regularization window.
    \end{corollary}
    \begin{proof}
        We first consider the constant predictor.
        Here, we have \[
            \abs{\predq_{i,e}^\predC(\theta, \bar\theta, f) - \predq_{i,e}^\predC(\theta',\bar\theta',f)} = \abs{q_e^f(\bar\theta) - q_e^f(\bar\theta')} \leq L_{q_e}\cdot \abs{\bar\theta - \bar\theta'} \leq L_{q_e} \cdot \norm{(\theta,\bar\theta) - (\theta',\bar\theta')}_\infty.
        \]
        For the perfect predictor we proceed analogously.

        To show the Lipschitz condition for the regularized linear predictor, we specify $\predq^\predRL_{i,e}(\theta,\bar\theta,f)\coloneqq q_e^f(\bar\theta)$ for predictions of the past, i.e. for $\theta < \bar\theta$.
        Because the pointwise maximum of a $L$-Lipschitz function with $0$ is again $L$-Lipschitz, we can omit the operator $(\emptyArg)^+$ in our considerations.
        For any $\theta,\bar\theta$ and $\theta',\bar\theta'$ we aim to show that \begin{align*}
            & \abs{q_e^f(\bar\theta) - q_e^f(\bar\theta')} \\
            & +\abs{
                \frac{q_e^f(\bar\theta)-q_e^f(\bar\theta-\delta)}{\delta}\cdot (\min\{\theta-\bar\theta, H\})^+
                -  \frac{q_e^f(\bar\theta')-q_e^f(\bar\theta'-\delta')}{\delta}\cdot (\min\{\theta'-\bar\theta', H\})^+
            }
        \end{align*}
        is upper bounded by $2 L_{q_e} (2+\nicefrac{H}{\delta})$.
        The first summand can be upper bounded by $L_{q_e}\cdot \norm{(\theta,\bar\theta)-(\theta',\bar\theta')}_\infty$.
        For the second summand $\alpha$, we abbreviate the formulas by introducing the notation $\alpha = \abs{\nicefrac{d}{\delta}\cdot m - \nicefrac{d'}{\delta}\cdot m'}$.
        Now, we compute \[
            \alpha \leq \frac{1}{\delta} \left( \abs{d\cdot m - d'\cdot m} + \abs{d'\cdot m - \cdot m'} \right)
            = \frac{1}{\delta} \left( m\cdot \abs{d-d'} +  \abs{d'}\cdot\abs{m-m'} \right).
        \]
        We find that $m \leq H$, $\abs{d - d'}\leq 2L_{q_e} \norm{(\theta,\bar\theta)-(\theta',\bar\theta')}$, $\abs{d'} \leq \delta\cdot L_{q_e}$ and $\abs{m-m'}\leq 2\norm{(\theta,\bar\theta)-(\theta',\bar\theta')}_\infty$.
        Simplifying yields\displayStyleForArxiv{\ensuremath{
                \alpha \leq 2 L_{q_e}\left( \nicefrac{H}{\delta} + 1 \right) \cdot \norm{(\theta,\bar\theta)-(\theta',\bar\theta')}.}}
        Summing up, we have shown that $\predq^\predRL$ is $(2L_{q_e}(2+\nicefrac{H}{\delta})$-Lipschitz.
    \end{proof}

    Note that if we add a time horizon to the ML predictors (such that the predicted queue remains constant after this horizon), and the mappings $\phi_{i,e}$ fulfill a Lipschitz-condition, then a similar result can be shown for the ML predictors.

 \section{Computational Study}

In the following computational study we compare the different predictors introduced in \Cref{sec:predictors} by measuring their performance in synthetic and real-world road networks.
As ML-based predictors we introduce both a linear regression based predictor $\predq^\predLR$ and a more advanced neural network based predictor~$\predq^\predNN$.

\subsection{Experiment Setup}

We run an experiment for each considered network.
The experiment consists of three phases:
In the first phase, we generate training data by computing $\delta_r$-routed DPE in which all commodities use the constant predictor.
Secondly, this data is used to train both the linear regression and the neural network based predictors.
In the third step, we evaluate the performance of all predictors by computing $\delta_r$-routed DPE in which all predictors are used side-by-side.

More specifically, we are given a network in which all commodities $i\in I$ use the constant predictor $\predq^{\text{C}}$.
Each commodity has a positive constant network inflow rate up to some common time $h$, after which all network inflow vanishes.
We sample this constant network inflow rate according to a normal distribution $\bar u_i\sim\mathcal N(\mu_i,\sigma_i^2)$ and set $u_i(\theta)\coloneqq \bar u_i$ for $\theta \in [0, h]$ and $u_i(\theta) = 0$ otherwise.
We compute several of these $\delta_r$-routed DPE according to \Cref{sec:outline-algo} with a fixed reroute-interval $\delta_r$ and randomly chosen network inflow rates, and use the resulting flows as training data.
\Cref{subsec:machine-learned-predictors} describes the machine-learned predictors in more detail.

For the evaluation step, we randomly pick one commodity $j\in I$ as the so-called \emph{focused commodity}.
For each introduced, causal predictor described in \Cref{predictors}, we add an additional commodity $i\in\{\predq^\predZ, \predq^\predC, \predq^\predL, \predq^\predRL, \predq^\predLR, \predq^\predNN \}$ using the corresponding predictor.
These extra commodities allow us to measure the predictors' performance:
They have the same source and sink as the focused commodity, and a very small constant inflow rate $\bar u_i$ up to time $h$.
This minimizes their influence on the flow of the original commodities.

With this setup, we compute a $\delta_r$-routed DPE.
To assess the performance, we monitor the average travel time of particles of the additional commodities:
The network outflow rate of a commodity $i$ is given by $o_i(\theta) \coloneqq \sum_{e\in\edgesEntering{t}} f_{i,e}^-(\theta) - \sum_{e\in\edgesLeaving{t}} f_{i,e}^+(\theta)$.
Taking the integral of $u_i(\psi) - o_i(\psi)$ over $[0, \phi]$ yields the flow of commodity $i$ inside the network at time $\phi$.
If we integrate this quantity over $\phi\in[0, H]$ with $H \geq h$, we obtain the \emph{total travel time} of particles of commodity $i$ up to time $H$:
\begin{align*}
    T^{\text{total}}_i
    &\coloneqq\int_0^H \int_0^\phi u_i(\psi) - o_{i}(\psi) \diff\psi \diff \phi
\end{align*}
Now, $T^{\text{avg}}_i\coloneqq  T_i^{\text{total}} / (h\cdot \bar{u}_i)$ denotes the \emph{average travel time}.
We compare these values against their optimum which can be computed as $T^{\text{avg}}_{i, \text{OPT}}\coloneqq \int_0^h\min\{H, l_{i,s}(\theta)\} - \theta \diff \theta/h$.
The \emph{slowdown} is defined as $T^\mathrm{avg}_i / T^\mathrm{avg}_{i,\mathrm{OPT}}-1$. 

\newcommand{\MAE}{\mathrm{MAE}}
We also monitor the accuracy of the predictor in hindsight:
Here, similar to the machine-learned predictor, we take samples of the predicted queue for each prediction time and take the mean absolute error compared to the actual queue.
More specifically, let $\delta_s$ denote the fixed step size of two consecutive interpolation points of the machine-learned predictor.
The \emph{mean absolute error} of predictor $\predq_i$ is defined as 
\[
    \MAE_i \coloneqq \frac{ \sum_{\bar\theta \in \bar\Theta} \sum_{j\in\firstN{k_f}} \sum_{e\in E} \abs{ \predq_{i,e}(\bar\theta + j\delta_s, \bar\theta, f) - q_e(\bar\theta + j\delta_s) }}{ \abs{\bar\Theta}\cdot k_f \cdot \abs{E} },
\]
where $\bar\Theta\coloneqq \set{j\delta_r | j\in\N_{0}, j\delta_r \leq H}$ is the set of times at which routes were computed.

We compute a total of $N$ $\delta_r$-routed DPE with randomly chosen commodities and network inflow rates, and aggregate the results to obtain reliable performance metrics.

\subsection{Data}

We conduct our experiment on several networks.
The first is a warm-up synthetic graph with $4$ nodes and $5$ edges as presented in \Cref{fig:synthetic-network}; it has a single commodity with a manually chosen demand $\mu_i$.
The second network called the \emph{Nguyen} network was taken from~\cite{Nguyen1982} augmented with capacities and edge transit times by~\cite{Han2019}.
Appropriate demand values were generated by hand.
The other graphs are real world road networks collected in~\citep{transportationnetworks}.
Thereof, we use the networks of the cities Sioux Falls and Anaheim.
These datasets come with edge attributes travel time $\tau_e$ and capacity $\nu_e$.
Moreover, they include static origin-destination demand values $\mu_i$.
We add versions of these networks with a single manually chosen commodity and a corresponding demand $\mu_i$. All manually chosen parameters can be found in \cite{GitHub2021}.
Details of each network are depicted in~\Cref{table:network-data}.

\begin{figure}
    \centering
    \begin{tikzpicture}
        \node[draw,circle](0)at(0,0) {$s$};
        \node[draw,circle](1)at(2,0) {$v$};
        \node[draw,circle](2)at(0,-2) {$t$};
        \node[draw,circle](3)at(2,-2) {$w$};
        
        \draw[thick,->](0) --node[above]{$(1,2)$} (1);
        \draw[thick,->](1) --node[right]{$(1,2)$} (3);
        \draw[thick,->](3) --node[above]{$(1,1)$} (2);
        \draw[thick,->](3) --node[above,sloped]{$(1,1)$} (0);
        \draw[thick,->](0) --node[left]{$(3,1)$} (2);
    \end{tikzpicture}
    \caption{A network with source $s$ and sink $t$. Edges are labeled with $(\tau_e, \nu_e)$.}\label{fig:synthetic-network}
\end{figure}

\newcommand{\avg}{\mathrm{avg}}
\begin{table}[ht]
    \caption{Attributes of the considered networks}
    \label{table:network-data}
    \centering
    \small
    \def\arraystretch{1.5}\setlength\tabcolsep{4pt}
    \begin{tabular}{c c c c c c c c c c }
        Network & {$\abs{E}$} & $\abs{V}$ & $\abs{I}$ & $\capa_{\avg}$ & $\transit_{\avg}$ & $\mu_{\avg}$ & $\sigma_\avg$
        \\
        \hline
        Synthetic & $5$ & $4$ & $1$ & $1{.}4$ & $1{.}4$ & $4$ & $0.5$
        \\
        Nguyen & $19$ & $13$ & $4$ & $50$ & $2{.}2$ & $100$ & $50$
        \\
        Sioux Falls I & $75$ & $24$ & $1$ & $10{,}247$ & $4{.}13$ & $8{,}000$ & $2{,}412$
        \\
        Sioux Falls II & $75$ & $24$ & $528$ & $10{,}247$ & $4{.}13$ & $682{.}95$ & $603$
\\
        Anaheim I & $914$ & $416$ & $1$ & $6{,}030$ & $0.88$ & $8{,}000$ & $900$
        \\
        Anaheim II & $914$ & $416$ & $1{.}406$ & $6{,}030$ & $0.88$ & $74.5$ & $900$
    \end{tabular}
\end{table}

Furthermore, we use the following parameters for computing the $\delta_r$-DPE in all networks:
The network inflow rates vanish at time $h=12$ and the computation horizon is set to $\Hcomp = 60$ which is large enough for all particles to reach the sink.
We set the reroute interval to $\delta_r = \nicefrac{1}{8}$.
The predictors $\predq^\predL$ and $\predq^\predRL$ use a prediction horizon of $H=20$, and $\predq^\predRL$ has a regularization window of $\delta = 1$.
The ML predictors $\predq^\predML$ have a step size of $\delta=1$ with $k_p = 20$ past and $k_f=20$ future samples.

\subsection{Machine-Learned Predictors}\label{subsec:machine-learned-predictors}

To assess the impact of different ML-based models in our setting, we train simple linear regression predictors as well as more advanced neural network predictors.
To obtain training data, we compute a number of $\delta_r$-routed DPE flows using the extension based algorithm of \Cref{sec:outline-algo}.
In these flows, all commodities use the constant predictor introduced in \Cref{predictors}.
This allows the model to estimate the progression of queues when agents follow instantaneously shortest paths.
Moreover, the predictors are trained separately for each network so that they can take local congestion effects into account.

\newcommand{\EIn}{E^{\mathrm{in}}}
\newcommand{\eIn}{e^{\mathrm{in}}}

\newcommand{\EOut}{E^{\mathrm{out}}}
\newcommand{\eOut}{e^{\mathrm{out}}}
The features for our models are the prediction time $\bar\theta$, $k_p=20$ observations of the past queue length and of the past edge load on a subset $\EIn = \{\eIn_1,\dots,\eIn_n\}$ of edges.
    This means, a single training sample can be written as the tuple
    \[
        X_i = \left(\begin{array}{ccc}
             & \bar\theta,  \\
             q_{\eIn_1}(\bar\theta - 0\cdot\delta),&\dots,&q_{\eIn_1}(\bar\theta-(k_p - 1)\delta),\\[.5em]
             & \vdots \\[.5em]
             q_{\eIn_n}(\bar\theta - 0\cdot\delta),&\dots,&q_{\eIn_n}(\bar\theta-(k_p - 1)\delta),\\
             L_{\eIn_1}(\bar\theta - 0\cdot\delta),&\dots,&L_{\eIn_1}(\bar\theta-(k_p - 1)\delta),\\[.5em]
             & \vdots \\[.5em]
             L_{\eIn_n}(\bar\theta - 0\cdot\delta),&\dots,&L_{\eIn_n}(\bar\theta-(k_p - 1)\delta)
        \end{array}\right).
    \]
The output of the model are $k_f=20$ samples of the predicted queue length of edges in $\EOut=\{\eOut_1,\dots,\eOut_m\}\subseteq E$.
    Thus, a corresponding label of the training sample can be written as \[
    Y_i = \left(\begin{array}{ccc}
             q_{\eOut_1}(\bar\theta + 1\cdot\delta),&\dots,&q_{\eOut_1}(\bar\theta + k_f\delta),\\[.5em]
             & \vdots \\[.5em]
             q_{\eOut_m}(\bar\theta + 1\cdot\delta),&\dots,&q_{\eOut_m}(\bar\theta + k_f\delta)
    \end{array}\right).
    \]
We implemented the following two variants:
For the first, a single model taking observations of all edges of the network as input is trained, i.e. $\EIn=\EOut=E$.
In the second, a model for each edge $e$ is trained that only uses observations of surrounding edges, i.e. $\EIn = N(e)$, $\EOut = \{ e\}$, where $N(e)$ consists of edges that are no further than 3 jumps away from $e$ on the undirected graph.

\subsubsection*{The Linear Regression Predictor}
As a first machine-learning method we employ a linear regression predictor $\predq^{\text{LR}}$.
In general, this method learns a weights matrix $W$ and a bias vector $b$ minimizing the error between $W X_i + b$ and $Y_i$.
We found that the Ridge regressor offered by the scikit-learn library \citep{scikit-learn} works well for our uses.
This regressor minimizes the linear squared error together with an $L_2$ regularization term.

\subsubsection*{The Neural Network Predictor}
To compare the capabilities of different machine-learning methods with different levels of complexity, we also develop a neural network based predictor $\predq^{\mathrm{NN}}$.
The input of the neural network is of the form $X_i$ as above.
The neural network consists of 4 densely connected layers. The first three layers have the same size as  $X_i$, only the last layer has the size of $Y_i$. 
The LeakyReLU function \[
\mathrm{LeakyReLU(x)} = \begin{cases}
    0.3\cdot x, &\text{for $x < 0$,}\\
    x, &\text{for $x\geq 0$,}
\end{cases}
\] acts as the activation function between the dense layers reducing the vanishing gradient problem.
An Adam optimizer minimizes the mean absolute error as the loss function.
Moreover, we apply an $L_2$ regularization term to the weights and biases of each layer; this helps to reduce over-fitting.

\subsection{Comparison of Predictors}\label{subsec:comparison-predictors}

  We first take a closer look at the synthetic network shown in~\Cref{fig:synthetic-network}.
    Here, a typical DPE, in which the single commodity with source $s$ and sink $t$ uses the constant predictor, shows the following behavior:
    Initially, the paths $st$ and $svwt$ are the predicted shortest $s$-$t$-paths both with a predicted traversal time of $3$.
    While there is a network inflow rate higher than $3$, queues at the edges $st$, $sv$ and $wt$ build up in such a way that both paths $st$ and $svwt$ have roughly the same predicted traversal time.
    Moreover, during that time, particles arriving at $w$ never leave their original path by taking edge $ws$.

    \begin{figure}
        \centering
        \subfloat[The slowdown compared to the minimum average travel time.]{\begin{tikzpicture}
        \begin{axis}
  [
  width=.45\textwidth,
  boxplot/draw direction = y,
  ylabel = {\begin{tabular}{c}Slowdown\\\small$T_i^{\mathrm{avg}} / T^{\mathrm{avg}}_{\text{OPT}} - 1$\end{tabular}},
log basis y={10},
  ymajorgrids=true,
  grid style=dashed,
  xtick = {1,2,3,4,5,6},
  xticklabels = {$\hat q^{\text{Z}}$,$\hat q^{\text{C}}$,$\hat q^{\text{L}}$,$\hat q^{\text{RL}}$,$\hat q^{\text{LR}}$,$\hat q^{\text{NN}}$},
  every axis plot/.append style = {fill, fill opacity = .1},
  ]
    \addplot + [
          mark = *,
          boxplot,
          solid,
          color=blue]
          table [row sep = \\, y index = 0] {
        data \\
        0.13020709918422746\\
0.11290776600496155\\
0.1301445408354549\\
0.17340896064424394\\
0.17067719615128296\\
0.11800483877448209\\
0.2218223962217103\\
0.12237155694081481\\
0.13953782315806662\\
0.1651830263234948\\
0.13958820675235928\\
0.18659572850544848\\
0.16867178283774442\\
0.15999530221189895\\
0.18274399237593708\\
0.10243720100178932\\
0.20124903129886573\\
0.20128811151184944\\
0.17848547613545818\\
0.19936954310181187\\
};
\addplot + [
          mark = *,
          boxplot,
          solid,
          color=red]
          table [row sep = \\, y index = 0] {
        data \\
        0.13628059627669709\\
0.11790420839602467\\
0.13621265705425922\\
0.18645621698197412\\
0.18379550045983573\\
0.12337944932033995\\
0.2548962775971293\\
0.12775344524451504\\
0.14656147902006933\\
0.1775233334032389\\
0.1458003088285611\\
0.20505858301778002\\
0.18191730959227814\\
0.16928985229617122\\
0.2004788081512059\\
0.10660227122371335\\
0.22586610014717157\\
0.22649001168208138\\
0.19323241260612\\
0.2238196702564763\\
};
\addplot + [
          mark = *,
          boxplot,
          solid,
          color={rgb,255:red,0; green,128; blue,0}]
          table [row sep = \\, y index = 0] {
        data \\
        0.04213622066057732\\
0.06272928923628585\\
0.04212289850479212\\
0.019250259142697956\\
0.019247661998660792\\
0.06603835831149607\\
0.019105453755163815\\
0.045531028052747\\
0.022644655375589462\\
0.01928263815125364\\
0.022656746201445532\\
0.019228322128087116\\
0.019257086078513064\\
0.019284347278901492\\
0.01923131689785662\\
0.06784161695185831\\
0.019174308233420145\\
0.01917009999849051\\
0.019202057869291522\\
0.019165980087990198\\
};
\addplot + [
          mark = *,
          boxplot,
          solid,
          color=orange]
          table [row sep = \\, y index = 0] {
        data \\
        0.019261953372375906\\
0.018532504678078654\\
0.019255119803297305\\
0.019250259142697956\\
0.019247661998660792\\
0.018518748773961535\\
0.019105453755163815\\
0.018565783700076333\\
0.019331854485810984\\
0.01928263815125364\\
0.01934159134372848\\
0.019228322128087116\\
0.019257086078513064\\
0.019284347278901492\\
0.01923131689785662\\
0.018899281820385028\\
0.019174308233420145\\
0.01917009999849051\\
0.019202057869291522\\
0.019165980087990198\\
};
\addplot + [
          mark = *,
          boxplot,
          solid,
          color=black]
          table [row sep = \\, y index = 0] {
        data \\
        0.0003210331352929696\\
0.0003247121432827793\\
0.0003207876711979196\\
0.0003187617576840829\\
0.0003187187520428125\\
0.0003223622704817686\\
0.0003163639499772497\\
0.0003227375189014392\\
0.00032011288106104097\\
0.00031929791614282443\\
0.00032049213424745204\\
0.0003183985063868189\\
0.00031887480374681587\\
0.00031932621729935384\\
0.0003184480962818981\\
0.0003281259178447016\\
0.000317504099543342\\
0.0003174344160947751\\
0.00031796360100511123\\
0.00031736619520361664\\
};
\addplot + [
          mark = *,
          boxplot,
          solid,
          color=black]
          table [row sep = \\, y index = 0] {
        data \\
        4.6557014417025755e-06\\
8.290262332177889e-06\\
4.593498021732145e-06\\
3.889386491984226e-06\\
3.888861779044461e-06\\
6.287272746652306e-06\\
3.8601295346829545e-06\\
5.9702030992259125e-06\\
3.905872293152868e-06\\
3.895928435193596e-06\\
3.970329710334397e-06\\
3.884954251853756e-06\\
3.890765824632325e-06\\
3.8962737574088635e-06\\
3.8855593387232545e-06\\
1.2994966987145418e-05\\
3.874041126561423e-06\\
3.873190880021582e-06\\
3.879647732407676e-06\\
3.872358480982996e-06\\
};
\end{axis}
    \end{tikzpicture}
     }
        \subfloat[The MAE of each predictor.]{\begin{tikzpicture}
        \begin{axis}
  [
  width=.45\textwidth,
  boxplot/draw direction = y,
  ylabel = {$\mathrm{MAE}_i$},
log basis y={10},
  ymajorgrids=true,
  grid style=dashed,
  xtick = {1,2,3,4,5,6},
  xticklabels = {$\hat q^{\text{Z}}$,$\hat q^{\text{C}}$,$\hat q^{\text{L}}$,$\hat q^{\text{RL}}$,$\hat q^{\text{LR}}$,$\hat q^{\text{NN}}$},
  every axis plot/.append style = {fill, fill opacity = .1},
  ]
    \addplot + [
          mark = *,
          boxplot,
          solid,
          color=blue]
          table [row sep = \\, y index = 0] {
        data \\
        0.8536464843330938\\
0.67124980790921\\
0.8586294365493023\\
1.3090782124165106\\
1.2823982958518148\\
0.7498497760366856\\
2.515101519340511\\
0.7664744712417925\\
0.9304222681924194\\
1.2000363732683748\\
0.9222030032381436\\
1.5321778784171098\\
1.2569344495884267\\
1.1310936633693565\\
1.4717906057128232\\
0.5517847316157544\\
1.8713954533146087\\
1.8847773923577094\\
1.4166058322238382\\
1.8458699366293956\\
};
\addplot + [
          mark = *,
          boxplot,
          solid,
          color=red]
          table [row sep = \\, y index = 0] {
        data \\
        0.6787880611990419\\
0.5589712823610565\\
0.6819891528015312\\
0.9222259403572871\\
0.9065381667564711\\
0.6082305823693692\\
1.4768978186316672\\
0.6199705208534042\\
0.7198320587389527\\
0.8601449357772901\\
0.7213975655401252\\
1.0342583119481075\\
0.8916169597370776\\
0.827399385249737\\
1.0030935043841147\\
0.48061815383163353\\
1.195201588524688\\
1.2006987902547601\\
0.9799011897634176\\
1.1844534337874768\\
};
\addplot + [
          mark = *,
          boxplot,
          solid,
          color={rgb,255:red,0; green,128; blue,0}]
          table [row sep = \\, y index = 0] {
        data \\
        0.8179608089717033\\
0.7751090015866476\\
0.8194953634966037\\
1.0963169583002046\\
1.0857795718736267\\
0.8101543795280233\\
1.580267310226405\\
0.8068623673241314\\
0.8831821435638835\\
1.0471875759881095\\
0.8534166312749998\\
1.210621549801536\\
1.075587003831598\\
0.9931639225745663\\
1.1900104426042377\\
0.736805001507105\\
1.3586366421974077\\
1.3627638412005298\\
1.1379337220987142\\
1.3515766640240074\\
};
\addplot + [
          mark = *,
          boxplot,
          solid,
          color=orange]
          table [row sep = \\, y index = 0] {
        data \\
        0.4679604065648147\\
0.43735438752647066\\
0.46820279649976204\\
0.6621592612899289\\
0.658647123568447\\
0.44753833709705204\\
0.9654726227471768\\
0.4675079283866881\\
0.4951087785159667\\
0.6279996839318579\\
0.47831834174475707\\
0.7394969837120768\\
0.6478935887826892\\
0.5840483811233416\\
0.7318761254690559\\
0.4344420054260374\\
0.8342552423065605\\
0.8421293416988107\\
0.6966325270492046\\
0.8437726195402172\\
};
\addplot + [
          mark = *,
          boxplot,
          solid,
          color=black]
          table [row sep = \\, y index = 0] {
        data \\
        0.11544136779159055\\
0.10446800920052554\\
0.11647567166393234\\
0.09570885959088671\\
0.09498923981986057\\
0.10239693486017878\\
0.4337907053274341\\
0.10232214109669645\\
0.11585201503647005\\
0.08950087158102753\\
0.11319848167852925\\
0.13154611586769774\\
0.09541915304121124\\
0.09402919180211393\\
0.12848975111774036\\
0.13217926566888363\\
0.22618639826737058\\
0.22747169483005678\\
0.12307356820796492\\
0.2226073326336186\\
};
\addplot + [
          mark = *,
          boxplot,
          solid,
          color=black]
          table [row sep = \\, y index = 0] {
        data \\
        0.032999160568149825\\
0.038770393999890365\\
0.032984528385235766\\
0.04558718769949727\\
0.045179900791742764\\
0.03579664973123922\\
0.1876082842360395\\
0.03664732621242379\\
0.0380960399508538\\
0.04200809284894225\\
0.036476865935272294\\
0.06472912344117035\\
0.045688388485352754\\
0.037983733555493\\
0.0615021900073582\\
0.052339599567302594\\
0.09531369435423577\\
0.09524708696967707\\
0.05261677964629379\\
0.09698051863580212\\
};
\end{axis}
    \end{tikzpicture}
     }
\caption{The experiment results of the synthetic network.}
        \label{fig:slowdown-synthetic-network}
    \end{figure}

    However, once the network inflow ends, the queues of $sv$ and $st$ start to decrease immediately.
    As soon as the queue of $st$ is small enough, particles arriving at $w$ start using both the path $wst$ and the direct connection $wt$.
    Eventually, all remaining particles arrive at $t$ without changing their route again.

    The return of some particles to $s$ from $w$ indicates that a more advanced predictor could anticipate the situation better and send these particles along edge $st$ in the first place.
    \Cref{fig:slowdown-synthetic-network} shows the slowdown and the MAE in $N=20$ evaluation runs in the experiment setup declared above, aggregated in two box plots.
    Here, we used the machine-learning variants with full-network observations for both the linear regression and the neural network predictors.

    The figure shows that the constant predictor actually has the highest slowdown in this scenario.
    Even the Zero-Predictor, which always sends particles at equal rates along both $st$ and $svwt$, has a slightly lower slowdown.
    While the linear and the regularized linear predictor already show a large improvement compared to the constant predictor and the Zero-Predictor, the two machine learned predictors win against those by another order of magnitude.

    In larger networks, these results are less pronounced: While the machine-learned predictors still clearly dominate the others, the difference between the other four predictors is less significant.
    In most experiments, $\predq^\predL$ performs the best out of these four.
    \Cref{fig:slowdown-anaheim-ii} shows the same box plots for the real-world network of Anaheim~II.

\begin{figure}
    \centering
    \subfloat[The slowdown compared to the minimum average travel time.]{\begin{tikzpicture}
        \begin{axis}
  [
  width=.45\textwidth,
  boxplot/draw direction = y,
  ylabel = {\begin{tabular}{c}Slowdown\\\small$T_i^{\mathrm{avg}} / T^{\mathrm{avg}}_{\text{OPT}} - 1$\end{tabular}},
log basis y={10},
  ymajorgrids=true,
  grid style=dashed,
  xtick = {1,2,3,4,5,6},
  xticklabels = {$\hat q^{\text{Z}}$,$\hat q^{\text{C}}$,$\hat q^{\text{L}}$,$\hat q^{\text{RL}}$,$\hat q^{\text{LR}}$,$\hat q^{\text{NN}}$},
  every axis plot/.append style = {fill, fill opacity = .1},
  ]
    \addplot + [
          mark = *,
          boxplot,
          solid,
          color=blue]
          table [row sep = \\, y index = 0] {
        data \\
        0.17781968320974895\\
0.0005100136380935538\\
0.2633356218519789\\
0.10575912147478372\\
0.017633214903465255\\
0.03388694061288944\\
0.016178711402149215\\
0.007133694100214916\\
0.00016068685326686705\\
0.1464470537588649\\
0.06673956900376465\\
0.1315290545692942\\
0.22936663701600168\\
0.06831524327146088\\
0.039975826750703325\\
0.1017754605420369\\
0.7297577264841462\\
0.26896828353408786\\
};
\addplot + [
          mark = *,
          boxplot,
          solid,
          color=red]
          table [row sep = \\, y index = 0] {
        data \\
        0.09594255736925494\\
0.004403097387548183\\
0.07578202349707985\\
0.017475780305879374\\
0.017633214903465255\\
0.05113486241968679\\
0.00989855973974163\\
0.0718350981736211\\
0.00016068685326686705\\
0.056329382956796126\\
0.08713679802141128\\
0.08475298814686427\\
0.28179253358805756\\
0.05288520582602718\\
0.023685907366111536\\
0.04093008490452288\\
0.08653167979252219\\
0.144264991393124\\
};
\addplot + [
          mark = *,
          boxplot,
          solid,
          color={rgb,255:red,0; green,128; blue,0}]
          table [row sep = \\, y index = 0] {
        data \\
        0.0869223255768441\\
0.007818874977271406\\
0.06458328583506767\\
0.007901575261088922\\
0.029234187122572752\\
0.04604585711703857\\
0.012282695118742248\\
0.10720649503263857\\
0.0006049222509729724\\
0.06758249986717813\\
0.05425306236591654\\
0.03230843941807571\\
0.11941031184425421\\
0.04401503673146001\\
0.022946021487935786\\
0.045535727816218285\\
0.05138587918055726\\
0.11919558321144375\\
};
\addplot + [
          mark = *,
          boxplot,
          solid,
          color=orange]
          table [row sep = \\, y index = 0] {
        data \\
        0.13074418024548362\\
0.0128868019174464\\
0.09470357784277894\\
0.01179265490015946\\
0.044610552661017655\\
0.05589331381099982\\
0.01314601937805504\\
0.12473699109819791\\
0.002309042356033908\\
0.08905285829138254\\
0.0589119726832652\\
0.05877611334999888\\
0.13518631447193452\\
0.05875001110066225\\
0.030696037758628725\\
0.06700881083559573\\
0.059195246841331706\\
0.11879958407037572\\
};
\addplot + [
          mark = *,
          boxplot,
          solid,
          color=black]
          table [row sep = \\, y index = 0] {
        data \\
        0.011757507726106953\\
0.0003938344219072132\\
0.005889475585431514\\
0.010425616397886639\\
0.00984461252166402\\
0.010524218078874714\\
0.0041957711987063995\\
0.028445401858654407\\
0.0008717437561820418\\
0.04554373085561925\\
0.018536249883767786\\
0.02466161695418423\\
0.0041949740479660935\\
0.04644608757800461\\
0.006525295220289884\\
0.014629103554052447\\
0.0010906780063486554\\
0.05564258300373859\\
};
\addplot + [
          mark = *,
          boxplot,
          solid,
          color=black]
          table [row sep = \\, y index = 0] {
        data \\
        0.010386937413597641\\
0.00042467723968431237\\
0.003799334532925913\\
0.0015415380184580219\\
0.007581846389772551\\
0.004426137517668405\\
0.0012942372063677876\\
0.02712624747347614\\
0.0003498388830003307\\
0.030539582533471554\\
0.008835796256772888\\
0.02972967815054961\\
0.0065723272504671915\\
0.01729865515524942\\
0.0032651227889970436\\
0.017030785337575516\\
0.00904110453823126\\
0.04563612506429027\\
};
\end{axis}
    \end{tikzpicture}
     }
    \subfloat[The MAE of each predictor.]{\begin{tikzpicture}
        \begin{axis}
  [
  width=.45\textwidth,
  boxplot/draw direction = y,
  ylabel = {$\mathrm{MAE}_i$},
log basis y={10},
  ymajorgrids=true,
  grid style=dashed,
  xtick = {1,2,3,4,5,6},
  xticklabels = {$\hat q^{\text{Z}}$,$\hat q^{\text{C}}$,$\hat q^{\text{L}}$,$\hat q^{\text{RL}}$,$\hat q^{\text{LR}}$,$\hat q^{\text{NN}}$},
  every axis plot/.append style = {fill, fill opacity = .1},
  ]
    \addplot + [
          mark = *,
          boxplot,
          solid,
          color=blue]
          table [row sep = \\, y index = 0] {
        data \\
        919.1340711221696\\
1192.382888689797\\
1148.0638421499034\\
1255.3290537760793\\
992.1547631745474\\
1376.7464572872777\\
1097.6842206821448\\
978.7446553207329\\
1347.3009159113683\\
1133.4911619783552\\
1098.9922396839345\\
1400.0834889435578\\
1327.4927477073816\\
1275.9335396967476\\
1277.898639207899\\
1140.5079419770277\\
1178.826151662629\\
1192.877104941247\\
1115.7058431444273\\
1211.2580289144423\\
};
\addplot + [
          mark = *,
          boxplot,
          solid,
          color=red]
          table [row sep = \\, y index = 0] {
        data \\
        1206.1351323267597\\
1458.2280138600818\\
1379.9712181649393\\
1465.4459731194822\\
1279.37978756016\\
1598.7037186503685\\
1363.2087294349926\\
1267.4674131303182\\
1648.5575156990128\\
1415.9110511665579\\
1355.780814505507\\
1625.070962619178\\
1510.4141959507974\\
1528.3397303809943\\
1532.7463698246693\\
1457.443978719537\\
1443.6798670354813\\
1429.799926805764\\
1390.4491329694472\\
1504.580176154177\\
};
\addplot + [
          mark = *,
          boxplot,
          solid,
          color={rgb,255:red,0; green,128; blue,0}]
          table [row sep = \\, y index = 0] {
        data \\
        2015.1674037124842\\
2211.3889960448378\\
2134.321393913565\\
2327.055212533803\\
2126.0294276933473\\
2464.272005299934\\
2170.327973825678\\
2093.6473783265665\\
2629.0898913731635\\
2226.747292959978\\
2224.901347934692\\
2485.9142720104273\\
2352.257120417488\\
2362.5728550084987\\
2382.7183929370394\\
2393.6655222016075\\
2278.987987885523\\
2270.493438801291\\
2201.3681700374495\\
2357.1378853558704\\
};
\addplot + [
          mark = *,
          boxplot,
          solid,
          color=orange]
          table [row sep = \\, y index = 0] {
        data \\
        2017.8433411980122\\
2200.8386988875804\\
2128.780001592873\\
2307.5978237400564\\
2113.127853997932\\
2457.5076989426116\\
2182.953524411993\\
2080.9048811167386\\
2600.9361026258975\\
2236.436176167925\\
2220.235616940418\\
2469.489766341417\\
2339.5741694427575\\
2382.148348769575\\
2373.306715304121\\
2366.5817390197076\\
2277.713638062496\\
2266.6400315028495\\
2202.102032024858\\
2349.4984308718826\\
};
\addplot + [
          mark = *,
          boxplot,
          solid,
          color=black]
          table [row sep = \\, y index = 0] {
        data \\
        461.2736011407242\\
435.7635231494216\\
488.77440810325305\\
506.6425345891765\\
460.97549723664304\\
512.8735731188387\\
449.07935880327705\\
460.84736395202316\\
562.9941028357232\\
446.57208194123166\\
481.0858219212051\\
505.7608213213085\\
493.0343064445415\\
520.8822987622547\\
484.0980460678456\\
468.07371314623896\\
471.9199626799321\\
495.1113146229137\\
449.1656506631201\\
475.81920269532594\\
};
\addplot + [
          mark = *,
          boxplot,
          solid,
          color=black]
          table [row sep = \\, y index = 0] {
        data \\
        232.78814476564153\\
228.60388888882486\\
238.92024196123765\\
276.11800500415865\\
235.12437774151442\\
282.8232282485791\\
241.89490446112166\\
234.83620600067997\\
314.1408622199085\\
229.05660414035813\\
255.0524566954065\\
286.5362443538317\\
264.1827691354901\\
274.93950354303865\\
264.1077648348916\\
254.0422214519287\\
249.27100449150208\\
257.7042437586488\\
232.79396800134066\\
262.80493867543913\\
};
\end{axis}
    \end{tikzpicture}
     }
\caption{The experiment results of the Anaheim II network.}
    \label{fig:slowdown-anaheim-ii}
\end{figure}

\begin{table}[tbhp]
    \caption{Results of the experiment in the considered networks.}
    \centering
    \small
    \def\arraystretch{1.5}\setlength\tabcolsep{4pt}
    \begin{tabular}{ r | c c c c c c | c c c }
        & \multicolumn{6}{c |}{Average Slowdown in $\%$}
        & \multicolumn{3}{c}{CPU Time}
        \\
        Network
        & $\predq^\predZ$ & $\predq^\predC$ & $\predq^\predL$ & $\predq^\predRL$  & $\predq^\predLR$ & $\predq^\predNN$
        & Train $\predq^\predLR$
        & Train $\predq^\predNN$
        & Evaluate
        
        \\
        \hline
        Synthetic  & 16.07 & 17.36 & 3.01 & 1.91 & 0.03 & 0.00 & 6s & 7m & 6m
        \\
        Nguyen & 2.02 & 1.21 & 2.56 & 1.36 & 0.27 & 0.35 & 14s & 10m & 10m
        \\
        Sioux Falls I & 7.78 & 7.16 & 4.12 & 4.61 & 0.73 & 0.06 & 9s & 5m & 46m
        \\
        Sioux Falls II & 5.81 & 6.18 & 4.25 & 4.99 & 0.26 & 0.09 & 1m & 1h45m & 47m
        \\
        Anaheim I & 22.00 & 22.11 & 14.67 & 16.22 & 0.50 & 0.23 & 6m & 1h35m & 6h31m
        \\
        Anaheim II & 13.36 & 6.68 & 5.11 & 6.48 & 1.67 & 1.25 & 33m & 18h35m & 41h49m
    \end{tabular}
    \label{table:network-results}
\end{table}

The results of the experiment are summarized in~\Cref{table:network-results}.
In the Anaheim II experiment, a total of 163 flows for training were computed to provide a total of 51{,}019 training samples for the machine-learned predictors.
In all other experiments, 500 training flows were computed yielding 198{,}500 training samples.
A training/validation split of $90\%/10\%$ was used.
Moreover, only for Anaheim I and Anaheim II the machine-learning variants with observations of only neighboring edges (at most three jumps away in the undirected graph) as inputs were used.
On average, this corresponds to $1.0\%$ of all edges. 
For all experiments, $N=20$ evaluation runs were carried out.
The CPU times listed were measured on an Intel\textsuperscript{\textregistered} Core\textsuperscript{\tiny TM} i7-1165G7 @ 2.80GHz while using all four cores.

\begin{table}[tbhp]
    \caption{
The approximation quality $\varepsilon_{\predq_i}$ of each predictor.}\label{table:approximation-quality}
    \centering
    \small
    \def\arraystretch{1.5}\setlength\tabcolsep{4pt}
    \begin{tabular}{ r | c c c c c c }
        & \multicolumn{6}{c}{Approximation quality $\varepsilon_{\predq_i}$}
        \\
        Network
        & $\predq^\predZ$ & $\predq^\predC$ & $\predq^\predL$ & $\predq^\predRL$  & $\predq^\predLR$ & $\predq^\predNN$
        
        \\
        \hline
        Synthetic      &  0.00 & 0.67 &  2.94 &  0.83 & 0.00 & 0.50
        \\
        Nguyen         &  0.00 &  0.74 &  25.52 &  5.47 & 0.68 & 0.63 
        \\
        Sioux Falls I  &  0.00 &  0.36 &  9.52 &  1.44 & 0.15 & 0.17
        \\
        Sioux Falls II &  0.00 &  0.61 &  24.23 &  8.06 & 0.20 & 0.41
        \\
        Anaheim I      &  0.00 & 0.58 & 11.83 & 6.09 & 0.51 & 0.54
        \\
        Anaheim II     &  0.00 &  0.65 &  6.57 & 11.25 & 0.52 & 0.46
    \end{tabular}
\end{table}

Table~\ref{table:approximation-quality} shows numerically obtained approximation qualities $\varepsilon_{\predq}$ for each predictor $\predq$ in each network where $\varepsilon_{\predq}$ is the minimal value such that \[
    \infl_{i,e}(\theta) > 0 \implies \predDel_{i,e}(\theta,\theta,f)\leq \varepsilon_{\predq_i}
\]
holds for all computed $\delta$-routed DPE flows $f$ of a particular network instance.
As the Zero-predictor never changes its predicted shortest routes, its approximation quality is perfect.
On the negative side, the linear and regularized linear predictors have worse approximation qualities.
This aligns with our theoretical findings as the linear predictor is not Lipschitz at all and the Lipschitz constant of the regularized linear predictor is quite high with $L=2 L_{q_e}(2+\nicefrac{H}{\delta})$.
Furthermore, we can see that for the constant predictor and the machine-learned predictors, a DPE can be approximated well with our method.

 \section{List of Notations}

\begin{longtable}{lp{12cm}}
	\textbf{Symbol}			& \textbf{Description} \\\hline\endhead
$\R$, $\R_{\geq0}$, $\R_{>0}$ & the set of all, non-negative and positive real numbers, respectively \\
        $\Lloc{p}(I, J)$ & the set of locally $p$-integrable functions $g:I\to J$ with $I,J\subseteq \R$ \\
        $C(A, B)$ & the set of continuous functions $g: A\to B$ \\
        $\RateFcts$ & the set of $g\in\Lloc{1}(\R,\R_{\geq 0})$ with $\restrict{g}{(-\infty,0)} \aequal 0$ \\
        $G=(V,E)$ & a directed graph with nodes $V$ and edges $E$ \\
        $\inEdges{v}, \outEdges{v}$ & the incoming and outgoing edges of node $v$ \\
        $\capa_e\in\R_{>0}$ & the capacity of an edge $e$ \\
        $\transit_e\in\R_{>0}$ & the (free-flow) transit time of an edge $e$ \\
        $I$ & the set of commodities \\
        $s_i,t_i\in V$ & the source and sink node of commodity $i$ \\
        $V_i\subseteq V$, $E_i\subseteq E$ & the nodes, edges lying on a $s_i$-$t_i$-path \\
        $u_i\in\RateFcts$ & the network inflow rate of commodity $i$ \\
        $f=(f^+, f^-)$ & a flow over time with $f^+, f^-\in\RateFcts^{I\times E}$ \\
        $\infl_{i,e}(\theta)$ & the inflow rate of particles of commodity $i$ into edge $e$ at time $\theta$ \\
        $\outfl_{i,e}(\theta)$ & the outflow rate of particles of commodity $i$ out of edge $e$ at time $\theta$ \\
        $F_{i,e}^+(\theta), F_{i,e}^-(\theta)$ & the cumulative in- and outflow until time $\theta$ \\
        $q_e(\theta)$ & the queue length of edge $e$ at time $\theta$ \\
        $\infl_e$, $\outfl_e$ & the aggregate in- and outflow rates $\infl_e\coloneqq \sum_{i\in I}\infl_{i,e},\outfl_e\coloneqq \sum_{i\in I}\outfl_{i,e}$ \\
        $b_{i,v}^-(\theta)$ & the inflow rate of particles of commodity $i$ into node $v$ \\
        $T_e(\theta)$ & the exit time when entering edge $e$ at time $\theta$,  $T_e(\theta)\coloneqq \theta+\nicefrac{q_e(\theta)}{\capa_e} + \transit_e$ \\
        $H\in\R$ & a time horizon \\
        $\predq_{i,e}(\theta,\bar\theta,f)$ & the predicted queue length of edge $e$ at time $\theta$ as predicted by commodity $i$ at time $\bar\theta$ given the (historical) flow $f$ \\
        $\predT_{i,e}(\theta,\bar\theta,f)$ & the predicted exit time of edge $e$ when entering at time $\theta$ \\
        $\predTP iP(\theta,\bar\theta,f)$ & the predicted exit time of path $P$ when entering at time $\theta$ \\
        $\predl_{i,v}(\theta,\bar\theta,f)$ & the predicted earliest arrival time at $t_i$ when departing at time $\theta$ in $v$ \\
        $\predDel_{i,e}(\theta,\bar\theta,f)$ & the predicted delay for taking edge $e$ at time $\theta$ \\
        $\predE_{i}(\theta,\bar\theta,f) \subseteq E_i$ & the set of $\bar\theta$-estimated active edges at time $\theta$
\end{longtable}
 
\paragraph{Funding.}
The research of the authors was funded by the Deutsche Forschungsgemeinschaft (DFG, German Research Foundation) - HA 8041/1-1 and HA 8041/4-1.

\paragraph{Acknowledgements.}
We would like to thank the three anonymous reviewers
for their constructive feedback during the reviewing process.

\printbibliography

\ifappendix
\fi

\end{document}